\documentclass[10pt, english,prd,twocolumn,superscriptaddress,nofootinbib,preprintnumbers,showpacs,floatfix]{revtex4-1}
\usepackage{amsmath}
\usepackage{amsfonts}
\usepackage{amssymb}
\usepackage{graphicx}
\usepackage{xcolor}
\definecolor{purple}{rgb}{1,0,1}
\definecolor{lime}{HTML}{A6CE39} % needs xcolor

\newcommand{\R}{\mathcal{R}}

\newcommand{\lsim}   {\mathrel{\mathop{\kern 0pt \rlap
  {\raise.2ex\hbox{$<$}}}
  \lower.9ex\hbox{\kern-.190em $\sim$}}}
\newcommand{\gsim}   {\mathrel{\mathop{\kern 0pt \rlap
  {\raise.2ex\hbox{$>$}}}
  \lower.9ex\hbox{\kern-.190em $\sim$}}}
\newcommand{\bw}{\begin{widetext}\begin{equation}}
\newcommand{\ew}{\end{equation}\end{widetext}}
\newcommand{\be}{\begin{equation}}
\newcommand{\ee}{\end{equation}}
\newcommand{\bea}{\begin{eqnarray}}
\newcommand{\eea}{\end{eqnarray}}

\newcommand{\nn}{\nonumber}

\begin{document}
%=================================================================
\title{Cosmic stringlike objects in hybrid metric-Palatini gravity}
%=================================================================
\author{Tiberiu Harko}
\email{tiberiu.harko@aira.astro.ro}
\affiliation{Astronomical Observatory, 19 Ciresilor Street, 400487 Cluj-Napoca, Romania,}
\affiliation{Faculty of Physics, Babes-Bolyai University, 1 Kogalniceanu Street,
400084 Cluj-Napoca, Romania}
\affiliation{School of Physics, Sun Yat-Sen University,  Xingang  Road, 510275 Guangzhou, People's
Republic of China}
%-----------------------------------------------------------------
\author{Francisco S. N. Lobo}
\email{fslobo@fc.ul.pt}
\affiliation{Instituto de Astrofísica e Ci\^encias do Espa\c{c}o, Faculdade de Ci\^encias da Universidade de Lisboa, Edif\'icio C8, Campo Grande, P-1749-016, Lisbon, Portugal}
%-----------------------------------------------------------------
\author{Hilberto M. R. da Silva}
\email{hilberto.silva@astro.up.pt}
\affiliation{Instituto de Astrof\'{i}sica e Ci\^encias do Espa\c{c}o, Universidade do Porto, CAUP, Rua das Estrelas, PT4150-762 Porto, Portugal, Centro de Astrof\'{i}sica da Universidade do Porto,
Rua das Estrelas, PT4150-762 Porto, Portugal}
%=================================================================
%%%%%%%%%%%%%%%%%%%%%%%%%%%%%%%%%%%%%%%%%%%%%%%%%%%%%%%%%%%%%%%%
\date{\LaTeX-ed \today}
%=================================================================
\begin{abstract}
We consider static and cylindrically symmetric interior string type solutions in the scalar-tensor representation of the hybrid metric-Palatini modified theory of gravity. As a first step in our study, we obtain the gravitational field equations and further simplify the analysis by imposing Lorentz invariance along the $t$ and $z$ axes, which reduces the number of unknown metric tensor components to a single function $W^2(r)$. In this case, the general solution of the field equations can be obtained, for an arbitrary form of the scalar field potential, in an exact closed parametric form, with the scalar field $\phi$ taken as a parameter. We consider in detail several exact solutions of the field equations, corresponding to a null and constant potential, and to a power-law potential of the form $V(\phi)=V_0\phi ^{3/4}$, in which the behaviors of the scalar field, of the metric tensor components and of the string tension can be described in a simple mathematical form. We also investigate the string models with exponential and Higgs type scalar field potentials by using numerical methods. In this way we obtain a large class of novel stable string-like solutions in the context of hybrid metric-Palatini gravity, in which the basic parameters, such as the scalar field, metric tensor components, and string tension, depend essentially on the initial values of the scalar field, and of its derivative, on the $r=0$ circular axis.
\end{abstract}
%=================================================================
\pacs{04.50.Kd, 04.20.Cv, 04.20.Fy}
%=================================================================
\maketitle
%%%%%%%%%%%%%%%%%%%%%%%%%%%%%%%%%%%%%%%%%%%%%%%%%%%%%%%%%%%%%
\def\d{{\mathrm{d}}}
%%%%%%%%%%%%%%%%%%%%%%%%%%%%%%%%%%%%%%%%%%%%%%%%%%%%%%%%%%%%%

%%%%%%%%%%%%%%%%%%%%%%%%%%%%%%%%%%%%%%%%%%%%%%%%%%%%%%%%%%%%%
\tableofcontents
%%%%%%%%%%%%%%%%%%%%%%%%%%%%%%%%%%%%%%%%%%%%%%%%%%%%%%%%%%%%%

%%%%%%%%%%%%%%%%%%%%%%%%%%%%%%%%%%%%%%%%%%%%%%%%%%%%%%%%%%%%%
\section{Introduction}
%%%%%%%%%%%%%%%%%%%%%%%%%%%%%%%%%%%%%%%%%%%%%%%%%%%%%%%%%%%%%

The formation of topological defects is a well studied physical process in the context of condensed matter, namely, metal crystallization \cite{Mermin:1979zz}, liquid crystals \cite{1975JMoSt..29..190J,1991Sci...251.1336C}, superfluid helium-3 \cite{1985PhRvL..55.1184S} and helium-4 \cite{1994Natur.368..315H}, and superconductivity \cite{Abrikosov:1956sx}. The formation of topological defects is a by-product of phase transitions and behind their formation lies a fundamental concept in physics, namely, spontaneous symmetry breaking (SSB).
Although we can distinguish first and second order phase transitions, the essential features of such a concept can be illustrated by a simple Goldstone model \cite{1962PhRv..127..965G,1994csot.book.....V, Math}.
Here, the physical system of a Higgs field exhibits a non-degenerate vacuum expectation value at $T>T_c$, but as the system is cooled, the minima of the potential becomes degenerate for $T<T_c$ and the field will ``roll'' to the new minima, where $T_c=\sqrt{6}\eta$ is the critical temperature in second order phase transitions, related to the energy breaking scale, $\eta$. The new vacuum state now does not exhibit the same invariance as the previous minimum and, hence, the symmetry is spontaneously broken.

From the standpoint of cosmology, the formation of topological defects is related with the symmetries shown by the Standard Model of Particle Physics. In fact, many Grand Unification Theories (GUTs) postulate that the universe, as it cooled, underwent a series of phase transitions associated with SSB, meaning that at sufficiently high temperatures there was invariance under a more general group of symmetries. Each of these phase transitions may have left behind a network of topological defects \cite{Kibble:1976sj}.
In fact, the Kibble and Zurek mechanism (KZM) \cite{Kibble:1976sj,Zurek:1985qw} describes the non-equilibrium dynamics and the formation of topological defects in a system which is driven through a continuous phase transition at finite rate.
When $T_c$ is reached, random fluctuations will dictate which of the minima state will be ``chosen'' by the field; regions of spacetime separated by a distance larger than the size of the particle horizon, will ``choose'' independent, but equivalent, states on the minima manifold. Indeed, the kind of defects we expect to be formed depend on the (non-trivial) topology of the minima manifold \cite{2016arXiv160702979A}.

In the case of a discrete symmetry breaking, whenever the vacuum manifold is disconnected, a domain wall is formed, which is a surface that separates two patches with different vacuum expectation values (VEV). If the vacuum manifold contains unshrinkable surfaces, the field might develop non-trivial configurations corresponding to point-like defects, known as monopoles. In this work we limit ourselves to the investigation of what are considered to be the most viable types of topological defects, which may have already formed in the early Universe, namely, cosmic strings \cite{1994csot.book.....V, Math}. These are line-like defects formed when the topology of the minima manifold is not simply-connected.

In field theoretical models strings can form once an axial symmetry is broken spontaneously. Strings can exist in the form of loops, or they can be infinitely long, spanning to the horizon. The equations of motion for two models of circular cosmic string loops with windings in a simply connected internal space were investigated numerically in \cite{LHJ}. The Kosambi-Cartan-Chern theory was used to analyze the Jacobi stability of the string equations and determine bounds on the physical parameters that ensure dynamical stability of the windings. One may also consider more exotic defects, composed of combinations of strings. For a more comprehensive discussion of the set of possible topological defects we refer the reader to \cite{1994csot.book.....V, Math}. The formation of a network of cosmic defects, and their symmetry breaking scale, is a key feature in many Grand Unified scenarios \cite{2003PhRvD..68j3514J}, and hence the search for the cosmological consequences of such defects is a key aspect for constraining different models.

Some defects tend to be inherently unstable \cite{1994csot.book.....V}, while domain walls and monopoles are either cosmologically catastrophic or severely constrained by current observations \cite{1991ApJ...373L..35N}. On the other hand, the presence of cosmic strings can have important cosmological consequences, such as, for example, in the case of the Cosmic Microwave Background (CMB) anisotropies  \cite{Ade:2013xla}, for small scale structure formation \cite{2002IJMPD..11...61W}, for the reionization history of the Universe \cite{2006PhRvD..74f3516O}, for Gamma Ray Bursts \cite{Cheng1,Cheng2}, for the  gravitational lensing observations \cite{2009PhRvL.103r1301T}, and for the understanding of the formation of the super-massive black holes in the early universe \cite{LH}, respectively.

Furthermore, the role of inflation on the survival of topological defects cannot be overstated, as defects formed too early would become diluted in the universe, which is useful in the case of domain walls or monopoles, but defects formed too late would become energetically dominant, changing drastically the standard cosmological model. However, it has been shown that strings are a by-product of several GUTs at the end of inflation \cite{2003PhRvD..68j3514J} and are stable topological defects, which make them good candidates for further analysis.
Due to the existence of a magnetic flux inside the string \cite{1973NuPhB..61...45N}, cosmic strings can either be infinite or form closed loops, which will oscillate and radiate energy via gravitational waves (GW), and thus decay. This radiation will cause a stochastic background in the GW spectra \cite{2018PhRvD..97j2002A}.

Even though most of the research on cosmic strings has been done in the framework of standard general relativity, the properties of cosmic strings have also been investigated in modified theories of gravity. String-like solutions have been found in $f(R)$ gravity  \cite{fR1,fR2,fR3}. Cylindrically symmetric string solutions with constant Ricci curvature have been derived in \cite{fR1}, and it was shown that there is only one solution for $R = 0$.  Families of
vacuum solutions for which $R = {\rm const} \neq 0$  were also found, representing $f(R)$ analogues of the Linet-Tian solution \cite{L,T}. In fact, the solution obtained in \cite{fR1} is a member of the  general Tian family of solutions in general relativity, and therefore it can describe the exterior of a cosmic string. Kasner-type static, cylindrically symmetric interior string solutions in the framework of $f\left(R,L_m\right)$ gravity \cite{Bertolami:2007gv, Harko:2010mv} were considered in \cite{Harko:2014axa}.
%The physical properties of the string are described by an anisotropic energy-momentum tensor satisfying the condition $T_t^t=T_z^z$, that is, the energy density of the string along the $z$-axis is equal to minus the string tension.
Gravitationally bound general relativistic strings consisting of a Bose-Einstein condensate matter that is described, in the Newtonian limit, by the zero temperature time-dependent nonlinear Schr\"{o}dinger equation (the Gross-Pitaevskii equation), with repulsive interparticle interactions were investigated in \cite{Bose}.

Cosmic strings have also been extensively explored in other extensions of general relativity, such as in scalar-tensor theories \cite{GundlachOrtiz,BarrosRomero,Guimares,BoisseauLinet,DahiaRomero,SenBB,Arazi,Gregory,SenB,Sen,Delice:2006gs}. An interesting aspect in these theories is the proof that the Vilenkin prescription in which an infinitely long straight static local gauge string satisfies the condition of the energy-momentum tensor $ T^t_t=T^z_z \neq 0$ and all the other components $T^\mu_{\nu} = 0$ \cite{Vilenkin}, is inconsistent in Brans-Dicke theory of gravity \cite{SenBB}. However, this inconsistency can be removed by including a cosmological constant \cite{Delice:2006gs}, or by considering a more general scalar-tensor theory \cite{SenB}.
This fact motivates investigating string type solutions in the scalar-tensor representation of the recently proposed hybrid metric-Palatini gravity \cite{Harko:2018ayt,Harko:2011nh,Capozziello:2012ny}, which is a modified theory of gravity that combines  the metric and Palatini formalisms, already introduced in the study of standard general relativity, to construct a new gravitational Lagrangian.

In fact, theories of gravity with a gravitational action consisting of more general combinations of curvature invariants than the traditional Einstein-Hilbert term have been in recent years a source of intense scrutiny \cite{Harko:2011kv,Sotiriou:2008rp,Capozziello:2011et,Nojiri:2010wj,Lobo:2008sg,Capozziello:2015lza,Harko:2018ayt}. Either by their ability to account for the late-time cosmic acceleration without dark energy \cite{Capozziello:2002rd}, or by the possibility of explaining the large scale dynamics of self-gravitating systems without the need for dark matter   \cite{Boehmer:2007kx,Bohmer:2007fh, Capozziello:2012qt,Capozziello:2013yha,Harko:2013xma}. The hybrid metric-Palatini theory is one of these cases \cite{Harko:2018ayt,Harko:2011nh,Capozziello:2012ny,
Capozziello:2013uya,Borowiec:2014wva,Capozziello:2015lza,Rosa:2020uoi,Rosa:2018jwp,Rosa:2017jld}. From a theoretical point of view the main advantage of hybrid metric-Palatini  gravity is that it is a viable gravity theory that includes elements of both Palatini and metric formalisms. A main success of the theory is the possibility of generating long-range forces that pass the classical local tests at the Solar System level of gravity. Another important advantage of the theory is that it admits an equivalent scalar-tensor representation, which greatly simplifies the analysis of the field equations, and the construction of their solutions. Vacuum solutions of the gravitational field equations in the hybrid metric-Palatini gravity were considered in \cite{Dan, Bron1, Bron2}.

In this work, we will analyze local gauge string solutions with a phenomenological energy momentum tensor, as prescribed by Vilenkin \cite{Vilenkin}, in the context of the hybrid gravitational theory. The general solution of the field equations can be obtained in an exact parametric form for arbitrary scalar field potentials. Several solutions of the field equations, obtained for different functional forms of the scalar field potential are considered in detail. In particular we consider the cases of the null and constant potentials, as well as the power-law potential of the form $V(\phi)=V_0\phi ^{3/4}$. For all these cases the solutions of the gravitational field equations can be represented in a simple mathematical form. Two important scalar field potentials are the exponential and the Higgs type scalar field potentials, which can be investigated only with the extensive use of numerical methods. As a result of our investigations we obtain several classes of novel stable string-like solutions in hybrid metric-Palatini gravity. An interesting property of these solutions is that the behavior of all physical and geometrical quantities describing  these string-like objects (scalar field, metric tensor components, and string tension), essentially depend on the initial values along the $r=0$ circular line of the scalar field, and of its derivative.

This work in organized in the following manner. In Sec.~\ref{sec:II}, we present the scalar-tensor representation of the hybrid metric-Palatini theory, by writing out the action and field equations  for a general static, cylindrically symmetric metric. This is followed by Sec.~\ref{sec:III}, where we  present the general solution of the field equations, for an arbitrary form of the scalar field potential, in an exact closed parametric form, with the scalar field $\phi$ taken as a parameter. In Sec.~\ref{sec:IV}, we consider in detail several exact and numerical solutions of the field equations, by choosing several interesting choices for the potential. Finally, we summarize and discuss our results in Sec.~\ref{sec:V}.

%\section{General formalism: hybrid metric-Palatini gravity}

%%%%%%%%%%%%%%%%%%%%%%%%%%%%%%%%%%%%%%%%%%%%%%%%%%%%%%%%%%%%%
\section{Cosmic string-like objects in hybrid metric-Palatini gravity}\label{sec:II}
%%%%%%%%%%%%%%%%%%%%%%%%%%%%%%%%%%%%%%%%%%%%%%%%%%%%%%%%%%%%%

In the present Section we will first briefly review some of the basic properties of the cosmic string and string-like objects. Then we introduce the action and the field equations of the hybrid metric-Palatini gravity theory, and we write down the system of equations describing cosmic strings in static cylindrical symmetry.

\subsection{Cosmic strings-a brief review}

 Cosmic strings naturally appear in spontaneously broken gauge theories,  which often exhibit stable topological defects. The simplest model that gives rise to a string solution is based on the matter Lagrangian \cite{Kibblereview, Ring}
\be\label{stringlag}
L_m=\left|\partial _{\mu}\Phi\right|^2-\frac{\lambda }{4}\left(|\Phi|^2-\bar{\eta} ^2\right)^2,
\ee
where $\lambda$ is a dimensionless coupling constant, while $\bar{\eta}$ is the vacuum expectation value of the field $\Phi$. The Lagrangian (\ref{stringlag}) is invariant under the gauge transformations $\phi (x)\rightarrow \phi (x)e^{i\alpha (x)}$, $A_{\mu}(x)\rightarrow A_{\mu}(x)-(1/e)\partial _{\mu}\alpha (x)$. A straight cosmic string oriented along the $z$-axis is a solution of the field equations of the form $\phi (t,z,\rho,\varphi)=\left(\eta /\sqrt{2}\right)f(\rho)e^{in\varphi}$, $A_{\mu}\left(t,z,\rho,\varphi\right)=-(n/e)h(\rho)\partial _{\mu}\phi$, where we have used cylindrical polar coordinates, and $n\in \mathbf{N}$ \cite{1973NuPhB..61...45N}. Generally the corresponding field equations cannot be solved analytically.  The model can also be extended to include gauge fields of the electromagnetic type \cite{1973NuPhB..61...45N}. In fact, the study of cosmic strings was pioneered in \cite{Kibble:1976sj}, and ever since it has become a popular subject of investigation. For a review of cosmic string and superstring properties see \cite{Kibblereview, Ring}.

Cosmic string configurations can also be obtained as solutions of the Einstein gravitational field equations. The string-like configurations are generally constructed by assuming  a cylindrically symmetric metric of the form \cite{Verbin1}
\be
ds^2=-N^2(r)dt^2+dr^2+L^2(r)d\theta ^2+K^2(r)dz^2.
\ee
 Such a metric with $K=N$ was used to give a complete classification of the string-like solutions of the gravitating Abelian Higgs model, where the functions $L$ and $N$ must satisfy the regularity conditions $L(0)=0$, $L'(0)=1$, $N(0)=1$, and $N'(0)=0$, respectively \cite{Verbin1}. Cosmic strings have the interesting property that around a straight, local cosmic string the spacetime is flat \cite{Kibblereview}.

For a cosmic string located along the $z$-axis, a solution of the Einstein field equations is given by \cite{Vilenkin}
\be\label{Vil1}
ds^2=-dt^2+dr^2+\left(1-8\pi G\mu \right)r^2d\theta ^2+dz^2.
\ee

The metric (\ref{Vil1}) was obtained  in the linear approximation of general relativity, by assuming that the matter energy-momentum tensor is given by $T_{\mu}^{\nu}(x,y)=\mu (x)\delta (x-a)\delta (y-b)\, {\rm diag}(1,0,0,1)$, where $\mu (x)$ is the linear energy-density of the string. Then, by representing the metric tensor as $g_{\mu \nu}=\eta _{\mu \nu}+h_{\mu \nu}$, where $\eta _{\mu \nu}$ is the Minkowski metric, and $h_{\mu \nu} \ll 1$, one obtains the gravitational field equations as given by $\left(\nabla ^2-\partial _t^2\right)h_{\mu \nu}=16\pi G\left(T_{\mu \nu}-(1/2)\eta _{\mu \nu}T\right)$, which must be considered together with the harmonic coordinate conditions $\partial _{\nu}\left(h_{\mu}^{\nu}-(1/2)\delta _{\mu}^{\nu}h\right)=0$ \cite{Vilenkin}. Then one obtains  (\ref{Vil1}) as a solution of the linearized field equations.

In terms of the modified azimuthal coordinate $\theta ' = \left(1 - 4G\mu \right)\theta$, ranging from 0 to $2\pi - 8 \pi G \mu$, the geometry described by the line element (\ref{Vil1}) is flat. This implies that the gravitational acceleration of massive objects towards the string is zero \cite{Vilenkin}. From a physical point of view this effect  is due to the equality of the tension and the energy per unit length, with the tension acting as a
negative source of the gravitational field. On the other hand it is important to point out that globally the spacetime is not flat, and it fundamentally has a conical form. Hence, the geometry of the massive string can be described as a conical singularity, having a deficit angle proportional to the mass density  $\mu$.

More generally, the metric for a string oriented along the $z$ axis and having an infinite length can be written as \cite{Meent}
\be\label{stringmet}
ds^2=-dt^2+dr^2+\left[1-\frac{\mu (r)}{2\pi}\right]^2r^2d\theta ^2+dz^2\,.
\ee
For this metric the Riemann curvature is zero everywhere, except along the $tz-$hyperplane, representing the string worldsheet.

Another interesting string configuration is given by the Barriola and Vilenkin string \cite{Barriola}, with metric
\bea
ds^2&=&-\left(1-8\pi \bar{\eta} ^2-\frac{2M}{r}\right)dt^2+\frac{dr^2}{\left(1-8\pi \bar{\eta} ^2-\frac{2M}{r}\right)}
	\nonumber  \\
&&\qquad +r^2\left(d\theta ^2+\sin ^2\theta d\varphi ^2\right).
\eea

This solution generalizes the matter Lagrangian (\ref{stringlag}) by considering a self-coupling scalar field triplet $\phi ^a$, $a = 1,2,3$, so that the matter action is given by $L_m=(1/2)\partial _{\mu} \phi ^a\partial ^{\mu}\phi ^a-\lambda\left(\phi ^a\phi ^a-\bar{\eta} ^2\right)^2$. To solve the gravitational field equations one uses the ansatz $\phi ^a=\eta f(r)x^a/r$.

In the framework of the Brans-Dicke theory static cylindrically symmetric solutions of the field equations have been obtained in \cite{Delice:2006gs}, for a gravitational action of the form
\begin{eqnarray}
S&=&\int d^4x
\sqrt{-g}\left[\phi(-R+2\Lambda)+\frac{\omega}{\phi}g^{\mu\nu}\partial_\mu\phi
\partial_\nu \phi\right]
	\nonumber\\&&
+  S_m[\Psi,g],
\end{eqnarray}
where $\Lambda$ is the cosmological constant.  For $\Lambda >0$ and a boost invariant cylindrically symmetric metric of the form (\ref{metrn}) (see below),  containing only a nontrivial metric tensor component $W^2(r)$, the solution of the field equations is given by \cite{Delice:2006gs}
\be
W(r)=A\sin ^{(\omega +1)/(\omega +2)}(\alpha r)\tan ^{\epsilon /(\omega +2)}\left(\frac{\alpha r}{2}\right),
\ee
\be
\phi (r)=A\sin ^{1/(\omega +2)}(\alpha r)\tan ^{-\epsilon /(\omega +2)}\left(\frac{\alpha r}{2}\right),
\ee
 where $\alpha =\sqrt{2\Lambda \left(2+\omega\right)}>0$, $\epsilon =\pm 1$, and $A$ is an integration constant. Similar solutions can be obtained in the case $\Lambda <0$, with the trigonometric functions replaced by the hyperbolic ones, so that
\be
W(r)=A\sinh ^{(\omega +1)/(\omega +2)}(\beta r)\tanh ^{\epsilon /(\omega +2)}\left(\frac{\beta r}{2}\right),
\ee
\be
\phi (r)=A\sinh ^{1/(\omega +2)}(\beta r)\tanh ^{-\epsilon /(\omega +2)}\left(\frac{\alpha r}{2}\right),
\ee
where $\beta =\sqrt{\left|2\Lambda \left(2+\omega\right)\right|}$. The regularity conditions $W(0)=0$ and $W'(0)=0$ on the axis for cylindrically symmetry of the string are satisfied if $A=2^{-\omega/(2\omega +4)}/\sqrt{(\omega +2)\Lambda}$ \cite{Delice:2006gs}. Hence for this choice of $A$ the solution is regular and free of any conical singularity. Moreover, the solutions with $\epsilon =1$ are smooth, regular, and free of any singularity on the axis $r=0$. Generally, the behavior of the solution and the regularity conditions depend on the adopted values of the model parameters $\left(A, \alpha, \beta\right)$.

%%%%%%%%%%%%%%%%%%%%%%%%%%%%%%%%%%%%%%%%%%%%%%%%%%%%%%%%%%%%%
\subsection{Action and field equations}
%%%%%%%%%%%%%%%%%%%%%%%%%%%%%%%%%%%%%%%%%%%%%%%%%%%%%%%%%%%%%

The action of hybrid metric-Palatini gravity is specified as \cite{Harko:2011nh,Capozziello:2012ny}
\begin{equation} \label{action}
S= \frac{1}{2\kappa^2}\int {\rm d}^4 x \sqrt{-g} \left[ R + f(\R)\right] +S_m \ ,
\end{equation}
where $S_m$ is the matter action, $\kappa^2\equiv 8\pi G$, $R$ is
the Einstein-Hilbert term, $\R  \equiv g^{\mu\nu}\R_{\mu\nu} $ is
the Palatini curvature, and $\R_{\mu\nu}$ is defined in terms of
an independent connection $\hat{\Gamma}^\alpha_{\mu\nu}$  as
\begin{equation}
\R_{\mu\nu} \equiv \hat{\Gamma}^\alpha_{\mu\nu ,\alpha} -
\hat{\Gamma}^\alpha_{\mu\alpha , \nu} +
\hat{\Gamma}^\alpha_{\alpha\lambda}\hat{\Gamma}^\lambda_{\mu\nu}
-\hat{\Gamma}^\alpha_{\mu\lambda}\hat{\Gamma}^\lambda_{\alpha\nu}\,.
\end{equation}

The action (\ref{action}) can be written in the scalar-tensor representation \cite{Harko:2011nh}, by the following action
\begin{eqnarray} \label{eq:S_scalar2}
S &=& \frac{1}{2\kappa^2}\int {\rm d}^4 x \sqrt{-g} \left[ (1+\phi)R +\frac{3}{2\phi}\partial_\mu \phi \partial^\mu \phi
-V(\phi)\right]
	\nonumber \\
&& \qquad \qquad  +S_m \ .
\end{eqnarray}
Performing the variation of the action with respect to the metric and the scalar field $\phi$ yields the field equations
\begin{eqnarray}\label{fieldeq}
    (1+\phi)G_{\mu\nu}=\kappa^2T_{\mu\nu}+\nabla_\mu\nabla_\nu\phi-\nabla_\alpha\nabla^\alpha\phi g_{\mu\nu}
	\nonumber  \\
   -\frac{3}{2\phi}\nabla_\mu\phi\nabla_\nu\phi +\frac{3}{4\phi}\nabla_\lambda\phi\nabla^\lambda\phi g_{\mu\nu}-\frac{1}{2}Vg_{\mu\nu}\,,
\end{eqnarray}
and
\begin{equation}\label{KG}
    -\nabla_\mu\nabla^\mu\phi+\frac{1}{2\phi}\partial_\mu\phi\partial^\mu\phi+\frac{\phi}{3} \left[ 2V-(1+\phi)V_{,\phi}\right]=\frac{\phi\kappa^2}{3}T \,,
\end{equation}
where $V_{,\phi}$ denotes the derivative of $V$ with respect to the scalar field. This equation of motion shows that, unlike in the Palatini case, the scalar field is dynamical and not affected by the microscopic instabilities found in Palatini models with infrared corrections \cite{Harko:2011nh}.

%%%%%%%%%%%%%%%%%%%%%%%%%%%%%%%%%%%%%%%%%%%%%%%%%%%%%%%%%%%%%
\subsection{Metric of a cosmic string-like object}\label{metrcs}
%%%%%%%%%%%%%%%%%%%%%%%%%%%%%%%%%%%%%%%%%%%%%%%%%%%%%%%%%%%%%

We now consider the specific case of a straight infinite cosmic string. Cosmic strings are part of the
class of line-like topological defects. This makes them different from the point-like monopoles and the membrane shaped domain walls. In any field theory the appearance of defects is related to the topology of the vacuum manifold \cite{Kibble:1976sj}.

In fact, as mentioned above, a crucial parameter in cosmic strings is the energy density, $\mu$ (which is usually represented as a dimensionless quantity $G\mu$) closely related to the energy scale of the symmetry breaking, $\eta\sim \sqrt{\mu}$; the tension of the string network, $G\mu$ is significantly constrained (for a detailed discussion see \cite{Ade:2013xla}), either by the CMB spectra ($G\mu<2.6\times 10^{-7}$) \cite{2010PhRvD..82b3521B}, gravitational lensing ($G\mu<10^{-9}$) \cite{2007PhRvD..76l3515M}, 21-cm observations ($G\mu<10^{-10}$) \cite{2008PhRvL.100i1302K} and, with the advent of LISA, it can be even more tightly constrained by the stochastic background GW spectra \cite{2016JCAP...04..001C}.

Throughout this paper, we will use Vilenkin's prescription \cite{Vilenkin}, given by
\begin{equation}\label{string}
    T^t_t=T^z_z=-\sigma(r) \,,
\end{equation}
where $\sigma$ is the string tension.

\subsubsection{Coordinates and geometry}

In the following we will consider an infinitely long, straight cylindrical string, whose matter content is described by the energy-momentum (\ref{string}). We assume that the geometry of the string has cylindrical symmetry, that is, rotational symmetry about the cylinder axis, together
with translational symmetry along the axis. Generally, a spacetime $V_4=(M, g)$  is called cylindrically symmetric if and only if it admits a $G_2$ on $S_2$ group of isometries that contain an axial symmetry, where $G_2$ is  the two-dimensional abelian Lie group (for an in depth discussion of the definition of cylindrical symmetry see \cite{Brondef} and references therein). If the spacetime admits two spacelike commuting Killing vector fields $\xi =\partial _z$  and $\eta =\partial _{\theta}$, both of them being mutually and  hypersurface-orthogonal, and $G_2$ acts orthogonally transitively, then the general cylindrically symmetric static metric takes the form  \cite{Steph}
\begin{equation}\label{metric}
    ds^2=-e^{2(K-U)}dt^2+e^{2(K-U)}dr^2+e^{-2U}W^2d\theta^2+e^{2U}dz^2,
\end{equation}
where $t$, $r$, $\theta$ and $z$ denote the time, radial, angular and axial cylindrical coordinates, respectively, and $K$, $U$ and $W$ are functions of $r$ alone.  The assumption of stationary axial symmetry requires for the metric to be invariant with respect to the transformations $x^0\rightarrow x^0+c_1$ (stationarity), $x^3\rightarrow x^3+c_2$ (axial symmetry), and $x^0\rightarrow -x^0$, $x^3\rightarrow -x^3$ (simultaneous reflection).

To admit an interpretation in terms of cylindrical symmetry, any cylindrically symmetric metric must have an axis. From a formal point of view the axis of symmetry can be defined as the set of fixed points of the map $\tau : T\times V_4\rightarrow V_4$, where $T$ is the one dimensional torus, that is, $W_2\equiv -x\in V_4$, $\tau _{\phi}(x)=x, \forall \phi \in T$, thus representing a two-dimensional surface $W_2$ \cite{Sen1,Sen2}.  In other words axial symmetry can be defined as an isometric $SO(2)$ mapping of space-time such that the set of fixed points forms a (regular) two-dimensional surface $W_2$, called the axis of rotation \cite{Steph}.  The condition of the existence of an axis is that the scalar products of the Killing vectors $\eta _a\eta ^a=e^{-2U(r)}W^2(r)$ vanishes on it, and the metric must be regular on this axis \cite{Steph}. For a metric with $U\equiv 0$ the condition for the existence of an axis is therefore $\left.W^2(r)\right|_{axis}=0$. If this condition is not satisfied, there is no axis for the system. Many of the known solutions of the Einstein field equations do not have an axis, or the axis is not regular \cite{Steph}. However, these metrics may be appropriate to describe the exterior field of the cylindrically symmetric systems containing a self gravitating mass distribution \cite{Steph}.

The solutions we will consider in the present investigation do not satisfy the condition for the existence of an axis, that is, they have the property that $\left.W^2(r)\right|_{axis}\neq 0$, and thus $W^2(r)$ does not vanish on any axis. Therefore a proper axis cannot be defined in this case, and the corresponding solutions of the field equations cannot be called cosmic strings in the usual sense. In the present paper we will call them string-like solutions. Moreover, in order to be able to provide a description of the obtained solutions in geometrical and physical terms we will call the $r=0$ line a circular line.

%%%%%%%%%%%%%%%%%%%%%%%%%%%%%%%%%%%%%%%%%%%%%%%%%%%%%%%%%%%%%
\subsection{Full field equations}
%%%%%%%%%%%%%%%%%%%%%%%%%%%%%%%%%%%%%%%%%%%%%%%%%%%%%%%%%%%%%

Taking into account the metric (\ref{metric}), then the field equation (\ref{fieldeq}) provides the following non-zero components
%\\$tt$-component
\begin{eqnarray}\label{tt}
    (1+\phi)\left(-U'^2+K' \frac{W'}{W}-\frac{W''}{W}\right)= \phi'' - \frac{3}{4\phi}\phi'^2
	\nonumber  \\
    - \left( K'-U' - \frac{W'}{W}  \right)\phi'   + \left( \kappa^2\sigma + \frac{1}{2}V \right) e^{2(K-U)} \,,
\end{eqnarray}
%
%$rr$-component
\begin{eqnarray}\label{rr}
    (1+\phi) \left(-U'^2+K' \frac{W'}{W} \right)=- \frac{3}{4\phi}\phi'^2
    	\nonumber  \\
    - \left( K'-U' +\frac{W'}{W} \right)\phi'   -  \frac{1}{2}V  e^{2(K-U)} \,,
\end{eqnarray}
%
%$\theta\theta$-component
\begin{equation}\label{thetatheta}
    (1+\phi) \left(U'^2+K'' \right)=-\phi''+\frac{3}{4\phi}\phi'^2 -U'\phi'    -\frac{1}{2}Ve^{2(K-U)}  \,,
\end{equation}
%
%$zz$-component
\begin{eqnarray}\label{zz}
&&    (1+\phi) \left(U'^2 + K'' -2U'' -2U'\frac{W'}{W}+\frac{W''}{W} \right)= -\phi''
	\nonumber  \\
&&    + \frac{3}{4\phi}\phi'^2
    + \left( U' - \frac{W'}{W}  \right)\phi'   -  \left( \kappa^2 \sigma + \frac{1}{2}V \right) e^{2(K-U)}.
\end{eqnarray}

Additionally, we can use Eq. (\ref{KG}) to determine the effective Klein-Gordon equation for the scalar field $\phi$:
\begin{eqnarray}\label{KG1}
    e^{-2(K-U)} \left(-\phi''-\frac{W'}{W}\phi'+\frac{\phi'^2}{2\phi} \right)&+&\frac{\phi}{3}\left[2V-(\phi+1)V_{,\phi}\right]
	\nonumber \\
    &+&\frac{2\phi\kappa^2\sigma}{3}=0 \,.
\end{eqnarray}

Since in this model the matter field couples minimally with curvature, it is possible to show that the energy conservation equation still holds, i.e.,
\begin{equation}\label{conservation}
    \nabla_\mu T^{\mu}{}_{\nu}=0
\end{equation}
which provides $K'\sigma=0$, and apart from the trivial vacuum solution, $\sigma=0$, this implies that $K'=0$. Thus,  we consider from now on that $e^K=1$, so that Eqs. (\ref{tt})--(\ref{zz}) simplify to the following relations
%\\$tt$-component
\begin{eqnarray}\label{tt1}
    (1+\phi)\left(-U'^2-\frac{W''}{W}\right)&=& \phi'' - \frac{3}{4\phi}\phi'^2
    + \left( U' + \frac{W'}{W}  \right)\phi'
	\nonumber  \\
 &&   + \left( \kappa^2\sigma + \frac{1}{2}V \right) e^{-2U} \,,
\end{eqnarray}
%\\$rr$-component
\begin{equation}\label{rr1}
    (1+\phi) U'^2 = \frac{3}{4\phi}\phi'^2
     +\left( -U' +\frac{W'}{W} \right)\phi'  +   \frac{1}{2}V  e^{-2U} \,,
\end{equation}
%\\$\theta\theta$-component
\begin{equation}\label{thetatheta1}
    (1+\phi) U'^2 =-\phi''+\frac{3}{4\phi}\phi'^2 -U'\phi'    -\frac{1}{2}Ve^{-2U} \,,
\end{equation}
%\\$zz$-component
\begin{eqnarray}\label{zz1}
&&    (1+\phi) \left(U'^2 -2U'' -2U'\frac{W'}{W}+\frac{W''}{W} \right)= -\phi'' + \frac{3}{4\phi}\phi'^2
	\nonumber \\
&&  \qquad \qquad   + \left( U' - \frac{W'}{W}  \right)\phi'   -  \left( \kappa^2 \sigma + \frac{1}{2}V \right) e^{-2U} \,,
\end{eqnarray}
respectively. Additionally, the effective Klein-Gordon equation for the scalar field $\phi$ reduces to
\begin{eqnarray}\label{KG2}
    e^{2U} \left(-\phi''-\frac{W'}{W}\phi'+\frac{\phi'^2}{2\phi} \right)&+&\frac{\phi}{3}\left[2V-(\phi+1)V_{,\phi}\right]
	\nonumber  \\
    &+&\frac{2\phi\kappa^2\sigma}{3}=0 \,.
\end{eqnarray}

%%%%%%%%%%%%%%%%%%%%%%%%%%%%%%%%%%%%%%%%%%%%%%%%%%%%%%%%%%%%%
\subsection{Field equations with boost invariance}
%%%%%%%%%%%%%%%%%%%%%%%%%%%%%%%%%%%%%%%%%%%%%%%%%%%%%%%%%%%%%

Note that local gauge strings preserve boost invariance along the $t$ and $z$ \cite{Vilenkin}, so that this requires $U=0$. Hence the only surviving non-trivial metric tensor component is $g_{\theta \theta}=W^2(r)$.  From a geometric point of view $W(r)$ is nothing but the radius of the coordinate
circles $r={\rm constant}$, $z={\rm constant}$, parameterized by the angle $\theta$. Since in this geometry the
circumference of a circle equals the value of $2\pi W$, in the following we will call the only remaining metric tensor component $W^2(r)$ a {\it circular radius}. On the other hand $W^2(r)$ also has the geometric meaning  of a length that may be counted from any zero point, with its value at $r=0$ not distinguished geometrically. Hence the metric of the cosmic string reduces to the form
\be\label{metrn}
ds^2=-dt^2+dr^2+W^2(r)d\theta^2+dz^2.
\ee

 Applying this symmetry, the gravitational field equations simplify considerably,
%\\$tt$-component
\begin{equation}\label{tt2}
(1+\phi)\left(-\frac{W''}{W}\right)= \phi'' - \frac{3}{4\phi}\phi'^2
+\frac{W'}{W}  \phi'   + \kappa^2\sigma + \frac{1}{2}V \,,
\end{equation}
%\\$rr$-component
\begin{equation}\label{rr2}
0= \frac{3}{4\phi}\phi'^2 +\frac{W'}{W}\phi'  +   \frac{1}{2}V  \,,
\end{equation}
%\\$\theta\theta$-component
\begin{equation}\label{thetatheta2}
0=-\phi''+\frac{3}{4\phi}\phi'^2     -\frac{1}{2}V  \,,
\end{equation}
%\\$zz$-component
\begin{equation}\label{zz2}
(1+\phi) \frac{W''}{W} = -\phi'' + \frac{3}{4\phi}\phi'^2
 - \frac{W'}{W}  \phi'   -   \kappa^2 \sigma - \frac{1}{2}V \,,
\end{equation}
where Eqs. (\ref{tt2}) and (\ref{zz2}) become redundant.

Combining Eqs. (\ref{rr2}) and (\ref{thetatheta2}) yields the following relation for the potential $V$:
\begin{equation}\label{simp2}
V=-\phi''-\frac{W'}{W}\phi'  \,,
\end{equation}
which substituting into the Klein-Gordon equation (\ref{KG2}), the latter reduces to:
\begin{equation}\label{KGsimp1}
V\left( 3+2\phi\right) -V_{,\phi}\phi\left( \phi+1\right) +2\kappa^2\sigma\phi + \frac{3\phi'^2}{2\phi}=0 \,.
\end{equation}
%which taking in to account the above expression takes the following form
%\begin{equation}
%\label{dynamicscalar}
%\left( -\phi''-\frac{W'}{W}\phi'\right) \phi-2\kappa^2\sigma\phi^3-\frac{W''}{W}\phi(\phi+1)+\frac{\phi'^2}{6}
%=0 \,.
%\end{equation}
Additionally, we can further deduce:
\begin{equation}\label{sigma1}
\kappa^2\sigma=\frac{1}{W}\left[(1+\phi)W'\right]' \,,
\end{equation}
and
\begin{equation}\label{this}
\frac{\left[(1+\phi)W\right]''}{W}=-(V+\kappa^2\sigma) \,.
\end{equation}

An important physical parameter characterizing the cosmic string properties is the mass per unit length of the string, which is defined as
\bea
m\left(R_s\right)&=&\int_0^{2\pi}{d\theta }\int_0^{R_s}{\sigma (r)W(r)dr}
	\nn \\
&=&2\pi \int_0^{R_s}{\sigma (r)W(r)dr},
\eea
where $R_s$ is the radius of the string-like object. In the following we will  interpret the string radius $R_s$ as representing  the maximum value of the coordinate $r$ at which the string density is non-zero, so that $\sigma (r) = 0$ for $r\geq R_s$.

\subsection{Regularity and asymptotic conditions}\label{reg}

A specific string-like solution of the gravitational field equations in cylindrical symmetry must satisfy a number of physical and geometrical requirements. In the case of self-gravitating systems we may go beyond the description of cosmic strings as simple conical singularities, and assume they are soliton-like structures \cite{Bronstring}. The soliton or string-like solutions of the field equations must satisfy a number of regularity and asymptotic conditions. The first of these conditions is the regularity along the string axis, which requires the absence of the conical singularity, and the finiteness of the algebraic curvature invariants, such as the Kretschmann scalar, $K=R_{\alpha \beta \gamma \delta}R^{\alpha \beta \gamma \delta}$. The condition of the finiteness of the Kretschmann invariants is satisfied if the $g_{00}$ and $g_{22}$ components of the metric tensor are finite on the axis. Equivalently, this condition can be formulated physically as the condition that all components of the energy-momentum tensor are finite.

The second condition for a string configuration obtained as a solution of the gravitational field equations is the requirement of its existence at  infinity, where the metric is either flat, or corresponds to the metric of a standard cosmic string configuration. This means first that the $g_{00}$ and the $g_{11}$ components of the metric tensor must tend at infinity to 1, $g_{00}\rightarrow 1$, $g_{11}\rightarrow 1$. Moreover, the curvature tensor must also vanish at infinity, and all the components of the energy-momentum tensor must decay rapidly.

Thirdly, we require that the total matter energy density per unit length of the string is finite, so that
\be\label{mint}
\int{T_0^0\sqrt{-^3g}d^3x}=\int{T_0^0e^{(K-U)}Wdrd\theta dz}<\infty.
\ee

In flat-space field theory this condition is used as a criterion for the field energy to be localized around the symmetry axis. One should point out that even the vacuum cylindrically symmetric solution of the Einstein gravitational field equations  in general has no regular asymptotic \cite{Bronstring}. From a physical point of view this can be explained by the infinite total mass of the infinitely long cylinder representing the string. The flat space-time metric, having a conical singularity on the axis \cite{Vilenkin}, is the only vacuum solution with a regular asymptotic \cite{Bronstring}. Static, linear, massless scalar fields, minimally coupled to gravity,  also cannot provide a regular asymptotic, with the mass integral (\ref{mint}) diverging at infinity. Another class of string-like are solutions that do not satisfy the asymptotic conditions are those obtained when the energy-momentum tensor behaves like a cosmological constant. Hence these solutions have asymptotics of cosmological nature. A few example of such models are those corresponding to closed models, like the Melvin magnetic universe, or those with de Sitter or anti-de Sitter behavior at infinity  \cite{Bronstring}.

%\section{Specific cosmic string solutions}
%%%%%%%%%%%%%%%%%%%%%%%%%%%%%%%%%%%%%%%%%%%%%%%%%%%%%%%%%%%%%
\section{General solution of the field equations}\label{sec:III}
%%%%%%%%%%%%%%%%%%%%%%%%%%%%%%%%%%%%%%%%%%%%%%%%%%%%%%%%%%%%%

In the present Section, we will consider the general solution of the field equations for a cosmic string in hybrid metric-Palatini gravity. It turns out that the system of gravitational equations describing a cosmic string can be solved analytically,  with the solution represented in an exact (closed) form, with all the geometric and physical quantities expressed in a parametric form, with the scalar field $\phi$ taken as a parameter. As an application of the obtained solution, in the next section, we will investigate the behavior of cosmic strings for several choices of the scalar field potential, including the cases of the constant potential, of the exponential potential, and of the Higgs-type potential, respectively.

By taking into account Eq.~(\ref{thetatheta2}), the field equations (\ref{tt2}) and (\ref{zz2}) reduce to the form
\be\label{de0}
\left(1+\phi\right)\frac{W''}{W}=-\frac{W'}{W}\phi '-\kappa ^2\sigma.
\ee
Equation (\ref{thetatheta2}) is independent of $W$ and, from a mathematical point of view, it represents a second order nonlinear differential equation. In order to solve it we first rescale the radial coordinate $r$ according to the transformation $r=\beta \xi$, where $\beta $ is an arbitrary length scale to be fixed from physical considerations. Hence Eq.~(\ref{thetatheta2}) takes the form
\be\label{de1}
\frac{d^2\phi}{d\xi ^2}-\frac{3}{4\phi}\left(\frac{d\phi}{d\xi}\right)^2+\frac{1}{2}\beta ^2V(\phi)=0.
\ee
In order to solve Eq.~(\ref{de1}) we introduce the transformations
\be
\frac{d\phi}{d\xi}=u, \quad \frac{d^2\phi}{d\xi ^2}=\frac{du}{d\xi}=\frac{du}{d\phi}\frac{d\phi}{d\xi}=u\frac{du}{d\phi}=\frac{1}{2}\frac{d}{d\phi}u^2,
\ee
and
\be
u^2=v,
\ee
respectively. Then Eq.~(\ref{de1}) becomes a first order linear differential equation of the form
\be
\frac{dv}{d\phi}-\frac{3}{2\phi}v+\beta ^2V(\phi)=0,
\ee
with the general solution given by
\be
v(\phi)=\phi ^{3/2}\left[C-\beta ^2\int{\phi ^{-3/2}V(\phi)d\phi}\right],
\ee
where $C$ is an arbitrary constant of integration. We immediately obtain
\be
u(\phi)=\phi ^{3/4}\sqrt{\left[C-\beta ^2\int{\phi ^{-3/2}V(\phi)d\phi}\right]},
\ee
and
\be\label{xi}
\xi +C _0=\int{\frac{\phi ^{-3/4}d\phi}{\sqrt{\left[C-\beta ^2\int{\phi ^{-3/2}V(\phi)d\phi}\right]}}},
\ee
respectively, where $C _0$ is an arbitrary constant of integration.

Equation~(\ref{rr2}) can be successively transformed as
\be
\frac{1}{W}\frac{dW}{d\xi}\frac{d\phi}{\d\xi}=-\frac{3}{4\phi}\left(\frac{d\phi}{d\xi}\right)^2-\frac{\beta ^2}{2}V(\phi),
\ee
and
\bea
\hspace{-0.5cm}&&\frac{1}{W}\frac{dW}{d\phi}=-\frac{3}{4\phi}-\frac{\beta ^2}{2}\frac{\phi ^{-3/2}V(\phi)}{\left[C-\beta ^2\int{\phi ^{-3/2}V(\phi)d\phi}\right]}
	\nonumber\\
\hspace{-0.5cm}&& \qquad = -\frac{3}{4\phi}+\frac{1}{2}\frac{d}{d\phi}\ln\left[C-\beta ^2\int{\phi ^{-3/2}V(\phi)d\phi}\right],
\eea
yielding
\be\label{W}
W(\phi)=W_0\phi ^{-3/4}\sqrt{C-\beta ^2\int{\phi ^{-3/2}V(\phi)d\phi}}\;,
\ee
where $W_0$ is an arbitrary constant of integration.

As a last step we need to obtain the expression of $\sigma$. Using Eq.~(\ref{rr2}), then Eq.~(\ref{de0}) can be rewritten as
\be\label{47}
\left(1+\phi\right)\frac{1}{W}\frac{d^2W}{d\xi^2}=\frac{3}{4\phi}\left(\frac{d\phi}{d\xi}\right)^2+\frac{1}{2}\beta ^2V(\phi)-\beta ^2\kappa ^2\sigma.
\ee
Taking into account the mathematical identities
\be
\frac{dW}{d\xi}=\frac{dW}{d\phi}\frac{d\phi}{d\xi}=\frac{dW}{d\phi}u,
\ee
\be
\frac{d^2W}{d\xi ^2}=\frac{d^2W}{d\phi ^2}v+\frac{1}{2}\frac{dW}{d\phi}\frac{dv}{d\phi},
\ee
respectively,  Eq. (\ref{47}) takes the form
\bea
(1+\phi)\left(\frac{1}{W}\frac{d^2W}{d\phi ^2}v+\frac{1}{2}\frac{1}{W}\frac{dW}{d\phi}\frac{dv}{d\phi}\right)
  \nn \\
=\frac{3}{4\phi}v+\frac{1}{2}\beta ^2V(\phi)-\beta ^2\kappa ^2\sigma.
\eea
Finally, after some simple calculations we obtain
\bea\label{sigma}
\kappa ^2\sigma (\phi)&=& \frac{1}{4 \phi} \left\{ \Big[2 \phi  (\phi +1) V'(\phi )+3 \sqrt{\phi } \int \frac{V(\phi )}{\phi^{3/2}} \, d\phi
	\right.
	\nn \\	
&& \left. -2 (2 \phi +3) V(\phi )\Big]-3 \left(C/\beta ^2\right) \sqrt{\phi } \right\}.
\eea

Equation (\ref{xi}), (\ref{W}) and (\ref{sigma}) give the complete solution of the field equations describing the geometry of a cosmic string in hybrid metric-Palatini gravity. The solution is obtained in a parametric form, with $\phi$ taken as a parameter. It also contains three arbitrary integration constants $C_0$, $C$, and $W_0$, respectively, which must be obtained from the initial or boundary conditions imposed on the cosmic string configuration.

As for the mass of the string, in the dimensionless variable $\xi$ it can be obtained as
\be
m(\xi_s)=2\pi \beta \int_0^{\xi _s}{\sigma (\xi)W(\xi)d\xi},
\ee
where $\xi _s=R_s/\beta$. We will also often consider  the condition $r\to\infty$, or, equivalently, $\xi\to \infty$ in the metric given by Eq.~(\ref{metrn}), which means that along the radial coordinate an infinite (very large) distance is to be covered.

%%%%%%%%%%%%%%%%%%%%%%%%%%%%%%%%%%%%%%%%%%%%%%%%%%%%%%%%%%%%%
\section{Specific cosmic string-like solutions}\label{sec:IV}
%%%%%%%%%%%%%%%%%%%%%%%%%%%%%%%%%%%%%%%%%%%%%%%%%%%%%%%%%%%%%

In the present section, we consider specific applications of the general solution of the field equations for a cosmic string in hybrid metric-Palatini gravity, outlined in the previous section. We will investigate the behavior of cosmic strings for several choices of the scalar field potential, including the cases of the constant potential, of the exponential potential, and of the Higgs type potential, respectively.

%%%%%%%%%%%%%%%%%%%%%%%%%%%%%%%%%%%%%%%%%%%%%%%%%%%%%%%%%%%%%
\subsection{Constant scalar field potential: $V=V_0$}
%%%%%%%%%%%%%%%%%%%%%%%%%%%%%%%%%%%%%%%%%%%%%%%%%%%%%%%%%%%%%

As a first example of a cosmic string model in the hybrid metric-Palatini modified theory of gravity, we will assume that the potential of the scalar field is a constant, $V(\phi)=V_0={\rm constant}$.

%%%%%%%%%%%%%%%%%%%%%%%%%%%%%%%%%%%%%%%%%%%%%%%%%%%%%%%%%%%%%
\subsubsection{The particular case $V=0$}
%%%%%%%%%%%%%%%%%%%%%%%%%%%%%%%%%%%%%%%%%%%%%%%%%%%%%%%%%%%%%

In the particular case $V=0$ the field equations describing the cosmic string configuration can be solved exactly. From Eq.~(\ref{xi}) we immediately obtain
\be\label{ss1}
\phi (\xi)=C \left(\xi-4C_0\right)^4.
\ee
The initial conditions $\phi (0)=\phi _0$ and $\phi '(0)=\phi_0'$ fix the constants $C_0$ and $C$ as
\be
C_0=-\frac{\phi _0}{\phi_0'},  \qquad  C=\frac{\phi_0'^{4}}{256\phi_0^3}.
\ee
For  $W$ we find
\be\label{ss2}
W^2(\xi)=\frac{w_0^2}{\left(\xi-4C_0\right)^6}.
\ee

 Hence from the above expression it follows that $W^2(\xi)$ does not satisfy the condition $W(0)=0$. On the $r=0$ circular line  the circular radius $W^2$ takes the finite value
 \begin{equation*}
 W^2(0)=W_0^2=\frac{w_0^2}{4096C_0^6}=\frac{w_0^2\phi _0^{\;\prime \;6}}{4096\phi _0^6},
 \end{equation*}
 a condition that fixes the integration constant $w_0^2$ as $w_0^2=4096W_0^2\phi _0^6/\phi _0^{\;\prime \;6}$. As for the energy density $\sigma $ of the string, it is given by
\be\label{ss3}
\kappa^2\sigma (\xi)=-\frac{12}{\left(\xi -4C_0\right)^2}.
\ee

Both the metric and the energy density are singular at $\xi =4C_0$. However, if  $C_0=-\phi _0/\phi_0^{\prime}<0$, implying that both $\phi _0$ and $\phi _0^{\prime}$ are positive, there is no infinite type singularity in the metric or energy density. The circular radius $W^2(\xi)$, the energy density $\sigma $, and the scalar field are monotonically increasing functions of $\xi$, with $\phi $ tending to infinity for $\xi \rightarrow \infty$.

As for the mass of the string, it is obtained as
\be
m\left(\xi _s\right)=24 \pi  \beta  W_0 \left[\frac{1}{1024
   C_0^4}-\frac{1}{4 (\xi _s-4
   C_0)^4}\right],
\ee
where $\xi _s$ is the string radius. If $\xi _s=4C_0$, the total mass of the string is (negative) infinite. On the other hand, for $C_0<0$, the string extends to infinity, but its mass is finite, taking the value
\be
\lim _{\xi _s\rightarrow \infty}m\left(\xi _s\right)=\frac{3 \pi  \beta  W_0}{128 C_0^4}=\frac{3 \pi  \beta  W_0 \phi _0^{\;\prime \;4}}{128 \phi _0^4}.
\ee

%%%%%%%%%%%%%%%%%%%%%%%%%%%%%%%%%%%%%%%%%%%%%%%%%%%%%%%%%%%%%
\subsubsection{The case $V=V_0\neq 0$}
%%%%%%%%%%%%%%%%%%%%%%%%%%%%%%%%%%%%%%%%%%%%%%%%%%%%%%%%%%%%%

We will proceed now to the general case of a constant potential, $V=V_0\neq 0$. Moreover, we will choose the scaling parameter of the radial coordinate $r$ so that $\beta ^2V_0=1$, giving $\beta =1/\sqrt{V_0}$, and $\xi =\sqrt{V_0}r$. Then the variation of the scalar field as a function of  $\xi $ is obtained from Eq.~(\ref{xi}) as
\be
\xi+C_0=\int{\frac{\phi ^{-3/4}d\phi}{\sqrt{C+2\phi ^{-1/2}}}},
\ee
giving
\be
C\left(\xi+C _0\right)=4 \sqrt[4]{\phi } \sqrt{C+\frac{2}{\sqrt{\phi }}}\;,
\ee
and
\be\label{field0}
\phi (\xi)=\frac{\left[C^2 \left(\xi +C _0\right)^2-32\right]^2}{256
   C^2},
\ee
respectively. The integration constants $C_0$ and $C$ must be determined from the initial conditions $\phi \left(\xi _0\right)=\phi _0$ and $\phi '\left(\xi _0\right)=\phi _0'$, respectively, and they are given by
\begin{equation}\label{const}
C_{0}=\pm \frac{2\phi _{0}-\phi _{0}^{\prime 2}}{
\phi _{0}^{3/2}}, \qquad C=\frac{4\phi _{0}\phi _{0}^{\prime }}{%
\phi _{0}^{\prime 2}-2\phi _{0}}-\xi _{0}.
\end{equation}

\begin{figure}[tbp]
 \centering
 \includegraphics[scale=0.65]{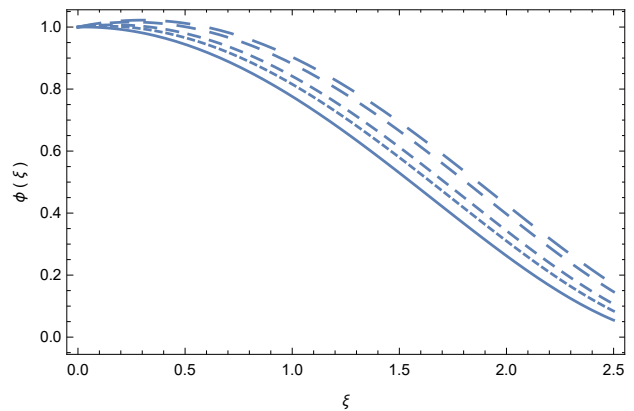}
 \caption{Variation of the scalar field of the cosmic string configuration in the presence of a constant potential for $\phi (0)=\phi_0=1$, and for different values of $\phi_0'$: $\phi_0'=0.012$ (solid curve), $\phi_0'=0.056$ (dotted curve), $\phi_0'=0.084$ (short dashed curve), $\phi_0'=0.126$ (dashed curve), and $\phi_0'=0.148$ (long dashed curve), respectively. }
 \label{fig1}
\end{figure}

The variation of the scalar field is represented in Fig.~\ref{fig1}.
For large values of the radial coordinate $r=\xi/\sqrt{V_0}$ the scalar field is a monotonically decreasing function of $\xi$, and, at large distances from the string, it reaches the value zero. The variation of $\phi$ is strongly dependent, from a quantitative point of view,  on the initial conditions for the field on the $r=0$ circular line. For large values of $\phi _0'$ and near the circular line, the scalar field is an increasing function of $\xi$,  and, after reaching a maximal value at a finite $r$, $\phi$ begins to decrease tending towards zero for very large values of $r$.

For  $W$ we obtain
\be
W(\phi)=W_0\phi ^{-3/4}\sqrt{C+2\phi ^{-1/2}},
\ee
or
\begin{equation}
W(\xi )=\frac{64 C^3 W_0 \left(\xi +C _0\right) }{\left[C^2 \left(\xi
   +C _0\right)^2-32\right]^2}.
\end{equation}

\begin{figure}[tbp]
 \centering
 \includegraphics[scale=0.65]{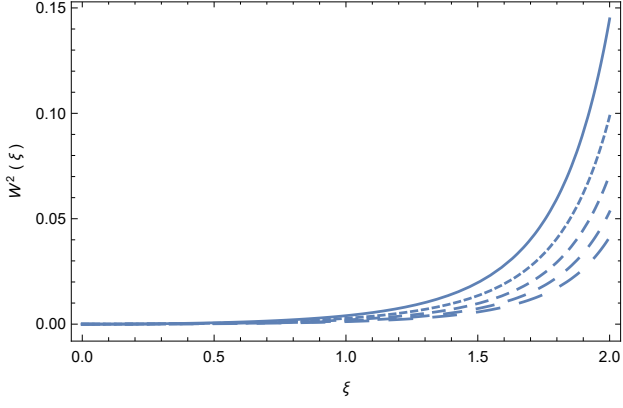}
 \caption{Variation of  $W^2(\xi)$ for the cosmic string configuration in the presence of a constant potential for $\phi \left(10^{-7}\right)=1$, $W\left(10^{-7}\right)=10^{-3}$, and for different values of $\phi_0'$: $\phi_0'=0.010$ (solid curve), $\phi_0'=0.012$ (dotted curve), $\phi_0'=0.014$ (short dashed curve), $\phi_0'=0.016$ (dashed curve), and $\phi_0'=0.018$ (long dashed curve), respectively. }
 \label{fig2}
\end{figure}

For $\xi =0$  we have
\begin{equation*}
W(0)=\frac{64C^3W_0C_0}{\left(C^2C_0^2-32\right)^2}.
 \end{equation*}
 The condition $W(0)=0$ would require to take $C_0=0$, which imposes the relation $2\phi _{0}=\phi _{0}^{\prime 2}$ between the initial values of the field and of its derivative on the string $r=0$ circular  line. But if this relation is satisfied, as can be seen immediately from the second of the Eqs.~(\ref{const}), the constant $C$ is undefined, and diverges for $\xi =0$. Therefore,  $W^2(\xi)$ is not defined for $r=0$.   The variation of  $W^2 (\xi)$ is represented, for $\phi \left(10^{-7}\right)=1$, $W\left(10^{-7}\right)=10^{-3}$, and for different values of $\phi _0'$, in Fig.~\ref{fig2}.

The circular radius $W^2(\xi)$ is divergent for $C^2 \left(\xi +C _0\right)^2-32=0$, which gives for the value of the singular point $\xi _{\infty}$ the expression
\be
\xi _{\infty}=\left(\phi _{0}^{\prime 2}-2\phi _{0}\right)\left[\frac{4\sqrt{2}}{4\phi _0\phi_0^{\prime}-\xi _0\left(\phi _{0}^{\prime 2}-2\phi _{0}\right)}\pm \frac{1}{\phi _0^{3/2}}\right].
\ee

The position of the singular point is essentially determined by the initial values of the scalar field and of its derivative near the $r=0$ circular line. At the metric singularity the scalar field vanishes, as one can see immediately from Eq.~(\ref{field0}). However, a different physical behavior is also possible, if near the origin the integration constants $C_0$ and $C$ satisfy the condition $C^2C_0^2 \gg 32$, or, equivalently, $\phi _0' \gg 2\phi _0$. In this case $W^2(\xi)$ can be approximated as
\be
W^2(\xi)\approx \frac{4096W_0^2}{C^2\left(\xi+C_0\right)^6}.
\ee
For $\xi \rightarrow \infty$, $W^2(\xi)\rightarrow 0$, and there are no infinity type singularities in the metric. However, a zero type singularity in the metric cannot be avoided even for this choice of the initial conditions.

\begin{figure}[tbp]
 \centering
 \includegraphics[scale=0.65]{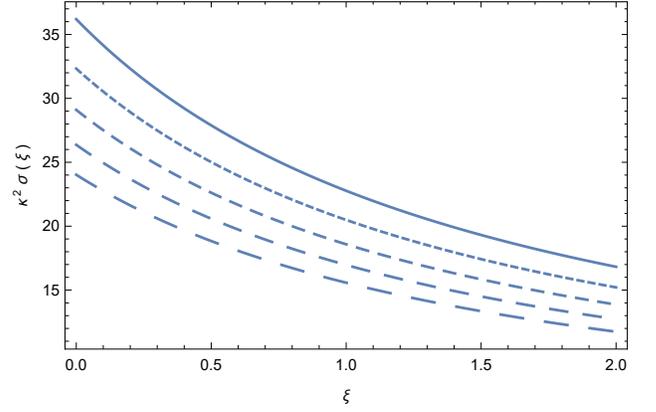}
 \caption{Variation of the energy density $\kappa ^2\sigma (\xi)$ of the cosmic string configuration in the presence of a constant potential for $\phi \left(0\right)=1$, and for different values of $\phi_0'$: $\phi_0'=1.272$ (solid curve), $\phi_0'=1.258$ (dotted curve), $\phi_0'=1.244$ (short dashed curve), $\phi_0'=1.230$ (dashed curve), and $\phi_0'=1.216$ (long dashed curve), respectively. }
 \label{fig3}
\end{figure}

As for $\sigma $, we easily find the expression
\be
\kappa ^2\sigma (\phi)=V_0\left( -\frac{3 C}{4 \sqrt{\phi }}- \frac{(\phi +3)}{\phi }\right) ,
\ee
or
\bea
\hspace{-0.4cm}&&\kappa ^2\sigma (\xi)=\frac{V_0}{\left[C^2 \left(\xi
   +C _0\right)^2-32\right]^2}
\Bigg\{C^2 \Big\{\left(\xi +C _0\right)^2  \times
 \nn \\
\hspace{-0.4cm}&& \left[-\left(C^2 \left((\xi
   +C _0)^2+12\right)-64\right)\right]-384\Big\}-1024 \Bigg\}.
\eea
The variation of $\sigma (\xi)$ is represented, for $\phi (0)=1$ and different values of $\phi'(0)$ in Fig.~\ref{fig3}.

In order to have positive values of $\sigma$ the integration constant $C$ must be negative, $C<0$, a condition that imposes some strong constraints on the initial values of the scalar field, and its derivative. The energy density of the string is a monotonically decreasing function of $\xi$, and, at least for the present choice of the initial conditions,  it does not have any singularities.

For the total mass of the string we obtain the expression
\bea
& m\left(\xi _s\right)=\frac{64\pi }{\kappa^2} \beta V_0 C W_0 \Bigg\{
   	%\nonumber\\
  - \frac{C_0^2 \left(C_0^2+24\right) C^4-64
   \left(C_0^2+8\right) C^2+1024}{\left(C_0^2
   C^2-32\right)^3}
   \nn \\
& +  \frac{C^4 (C_0+\xi _s)^2
   \left[\left(C_0+\xi _s\right)^2+24\right]-64 C^2 \left[\left(C_0+\xi _s\right)^2+8\right]+1024}{\left[C^2 (C_0+\xi _s)^2-32\right]^3}\Bigg\}.
\eea

The mass is divergent for $C_0^2C^2=32$, and for $C^2 (C_0+\xi _s)^2\rightarrow 32$. If the string radius tends to infinity, the mass of the string is finite, and it is given by
\bea
\hspace{-0.5cm}&&\lim _{\xi _s \rightarrow \infty}m\left(\xi _s\right)=\frac{64 \pi  \beta  C W_0}{\left(32-C_0^2
   C^2\right)^3} \times
	\nn \\
\hspace{-0.5cm} && \quad \times  \left[C_0^2 \left(C_0^2+24\right)
   C^4-64 \left(C_0^2+8\right) C^2+1024\right],
\eea
where we have assumed that $C_0^2C^2<32$.

%%%%%%%%%%%%%%%%%%%%%%%%%%%%%%%%%%%%%%%%%%%%%%%%%%%%%%%%%%%%%
\subsection{Power law potential: $V(\phi)=V_0\phi ^{3/4}$}
%%%%%%%%%%%%%%%%%%%%%%%%%%%%%%%%%%%%%%%%%%%%%%%%%%%%%%%%%%%%%

The gravitational field equations describing a cosmic string in hybrid metric-Palatini gravity also admit another exact solution, corresponding to the power law type scalar field potential $V(\phi)=V_0\phi ^{3/4}$. We rescale the radial coordinate $r$ by imposing the condition $\beta ^2V_0=1$, which gives $r=\xi/\sqrt{V_0}$. With these choices from Eq.~(\ref{xi}) we obtain explicitly the scalar field as a function of $\xi$, given by
\be
\phi (\xi )=\frac{\left( \xi ^{2}\phi _{0}^{3/4}\pm 2\xi \text{$%
\phi $}_{0}^{\prime }\pm 8\phi _{0}\right) ^{4}}{4096\phi
_{0}^{3}},
\ee
where we have used the usual initial conditions $\phi (0)=\phi _0$ and $\phi '(0)=\phi _0^{\prime }$, respectively. For the circular radius $W$ we obtain
\be
W(\xi)=\frac{W_0}{\left(\xi ^2 \phi _0^{3/4}\pm 2 \xi  \phi_0^{\prime}\pm 8 \phi
   _0\right)^3 \sqrt{2 \xi \pm 2 \phi _0^{\prime}/\phi _0^{3/4}}}.
\ee
where $W_0$ is an arbitrary constant of integration. For  $\xi =0$, we obtain $W^2(0)=\pm W_0^2/524288 \phi _0^{21/4} \phi _0^{\prime}$. Since  $W^2$ must be positive for all $\xi \geq 0$, it follows that the physical solution for the string configuration is the one with the positive sign. Hence in the case of the $V(\phi)=V_0\phi ^{3/4}$ potential, the solutions of the field equations describing a cosmic string in hybrid metric-Palatini gravity are
\bea
\phi (\xi)&=&\frac{\left( \xi ^{2}\phi _{0}^{3/4}+ 2\xi \text{$%
\phi $}_{0}^{\prime }+ 8\phi _{0}\right) ^{4}}{4096\phi
_{0}^{3}},
   \nn	\\
W^2(\xi)&=&\frac{W_0^2}{\left(\xi ^2 \phi _0^{3/4}+2 \xi  \phi_0^{\prime}+ 8 \phi
   _0\right)^6 \left(2 \xi + 2 \phi _0^{\prime}/\phi _0^{3/4}\right)}, \nn
\eea
respectively, with $W_0^2=524288W^2(0) \phi _0^{21/4} \phi _0^{\prime}$, a condition that implies $\phi _0>0$ and  $\phi _0^{\prime}>0$. For the string tension we obtain the expressions
\be
\kappa^2\sigma (\phi)=V_0\frac{-6 C-5(\phi -3) \sqrt[4]{\phi }}{8 \sqrt{\phi }},
\ee
and

\bea
\kappa^2\sigma (\xi)&=&\frac{V_0 \phi _0^{3/4}}{\left(\xi ^2 \phi _0^{3/4}+2 \xi  \phi_0'+8 \phi _0\right)^2}
 \Big\{-48C \phi _0^{3/4}
	\nn \\
   &&\quad -5 \left(\xi ^2
   \phi _0^{3/4}+2 \xi  \phi_0'+8 \phi _0\right) \times
   \nn\\
   && \quad \times \left[\frac{(\xi ^2 \phi _0^{3/4}+2 \xi \phi _0'+8 \phi _0)^4}{4096 \phi _0^3}-3\right]\Bigg\},
\eea
respectively.

%\begin{widetext}
%\be
%\kappa^2\sigma (\xi)=V_0\frac{\phi _0^{3/4} \left\{-48C \phi _0^{3/4}-5 \left(\xi ^2
%   \phi _0^{3/4}+2 \xi  \phi_0'+8 \phi _0\right)
%   \left[\frac{\left(\xi ^2 \phi _0^{3/4}+2 \xi \phi _0'+8 \phi _0\right)^4}{4096 \phi _0^3}-3\right]\right\}}
%{\left(\xi ^2
%   \phi _0^{3/4}+2 \xi  \phi_0'+8 \phi _0\right)^2},
%\ee
%\end{widetext}

In this case, the scalar field is a monotonically increasing function of the distance $\xi$  from the string, and tends to infinity for $\xi \rightarrow \infty$. On the other hand  $W^2(\xi)$ decreases monotonically from a finite value at $\xi =0$ to zero at infinity.  For $\xi =0$, the string tension takes the finite value
\begin{equation*}
\sigma (0)=\frac{V_0\left[-48 C \phi _0^{3/4}-40 \left(\phi _0-3\right) \phi _0\right]}{64
   \phi _0^{5/4}},
   \end{equation*}
while $\lim _{\xi \rightarrow \infty}\sigma (\xi)=-\infty$, indicating that $\sigma $ is a monotonically decreasing function of $\xi$. In the first order of approximation we obtain for the mass of the string of radius $\xi _s$ the expression
 \bea
  && m\left(\xi _s\right)=\frac{\pi  \beta  W_0  \xi _s}{8192 \phi_0^{33/8} \phi _0^{\;\prime \;3/2}}\times \nonumber\\
 && \Big\{6 C \xi _s \phi_0^{7/4}+15 \xi _s \phi _0^{\;\prime \;2} \left(C-2 \sqrt[4]{\phi_0}\right)-4 \phi_0 \phi_0' \times
		\nn \\
    && \quad \left[6 C+5 \left(\phi_0-3\right)
   \sqrt[4]{\phi_0}\right]+5 \xi _s \left(\phi_0-3\right) \phi_0^2\Big\}
   {}.
   \eea
In this approximation the mass is monotonically increasing with the string radius.

%\begin{widetext}
%   \be
%   m=\frac{\pi  \beta  W_0  \xi _s\left\{6 C \xi _s \phi_0^{7/4}+15 \xi _s \phi _0^{\;\prime \;2} \left(C-2 \sqrt[4]
%	{\phi_0}\right)-4 \phi_0 \phi_0' \left[6 C+5 \left(\phi_0-3\right)
%   \sqrt[4]{\phi_0}\right]+5 \xi _s \left(\phi_0-3\right) \phi_0^2\right\}}{8192 \phi_0^{33/8} \phi _0^{\;\prime \;
%	3/2}}.
%   \ee
%\end{widetext}

%%%%%%%%%%%%%%%%%%%%%%%%%%%%%%%%%%%%%%%%%%%%%%%%%%%%%%%%%%%%%
\subsection{Exponential potential: $V(\phi)=V_0e^{-\lambda \phi}$ }
%%%%%%%%%%%%%%%%%%%%%%%%%%%%%%%%%%%%%%%%%%%%%%%%%%%%%%%%%%%%%

As a second example of a string type configuration, we will consider the configuration generated by an exponential type potential, with $V(\phi)=V_0e^{-\lambda \phi}$, where $V_0$ and $\lambda >0$ are constants. The solutions of the gravitational field equations for different scalar field models with exponential potentials have been intensively investigated in the recent physical literature, including the cases of both homogeneous and inhomogeneous scalar fields \cite{exp1,exp2,exp3,exp4,exp5,exp6,exp7,exp8}. In four-dimensional effective Kaluza-Klein or string-type theories
 an exponential potential is generated from the compactification of the higher dimensions \cite{exp9}. Due to the curvature of the internal spaces or to the interaction with form fields on the
internal spaces, the moduli fields may acquire exponential type potentials. Non-perturbative effects such as gaugino condensation can also lead to exponential type potentials for scalar fields \cite{exp10}.

In the case of the exponential potential Eq.~(\ref{xi}) giving the scalar field-radial coordinate dependence becomes
\begin{eqnarray}\label{xiexp}
\hspace{-0.7cm}&\xi +C_{0}=\int {\frac{\phi ^{-3/4}d\phi }{\sqrt{C+2\sqrt{\pi }\beta ^{2}V_{0}%
\sqrt{\lambda}\,\text{erf}\left( \sqrt{\lambda \phi }\right) +2\beta
^{2}V_{0}e^{-\lambda \phi }/\sqrt{\phi }}}} ,
\end{eqnarray}
where $\text{erf}(x)$ is the error function, and cannot be represented in a closed form, therefore we will use a numerical approach to solve the field equations. We rescale first the scalar field  so that $\phi =\Phi/\lambda$, and we choose the scaling parameter $\beta $ of the radial coordinate as $\beta =\sqrt{2/V_0\lambda}$. Then Eq.~(\ref{de1}), which gives the variation of the scalar field, takes the form
\be
\frac{d^2\Phi}{d\xi ^2}-\frac{3}{4\Phi}\left(\frac{d\Phi}{d\xi}\right)^2+e^{-\Phi}=0.
\ee

The variation of  $W^2$ can be obtained from the equation
\be
\frac{1}{W}\frac{dW}{d\xi}\frac{d\Phi}{\d\xi}=-\frac{3}{4\Phi}\left(\frac{d\Phi}{d\xi}\right)^2-e^{-\Phi}.
\ee

\begin{figure}[tbp]
 \centering
 \includegraphics[scale=0.65]{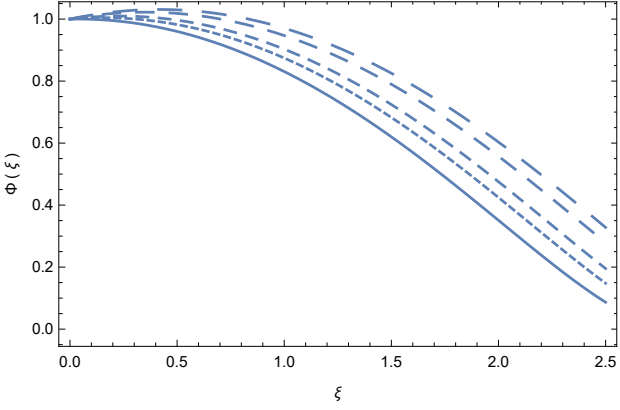}
 \caption{Variation of the scalar field of the cosmic string configuration in the presence of an exponential potential $V(\Phi)=e^{-\Phi}$ for $\Phi (0)=\Phi_0=1$, and for different values of $\Phi_0'$: $\Phi_0'=0.012$ (solid curve), $\Phi_0'=0.056$ (dotted curve), $\Phi_0'=0.084$ (short dashed curve), $\Phi_0'=0.126$ (dashed curve), and $\Phi_0'=0.148$ (long dashed curve), respectively.}
 \label{fig4}
\end{figure}

The behavior of the scalar field with exponential type potential is represented in Fig.~\ref{fig4}. For the sake of comparison we have chosen the same initial values for the field $\Phi$ and for its derivative as in the case of the constant potential.
\begin{figure}[tbp]
 \centering
 \includegraphics[scale=0.65]{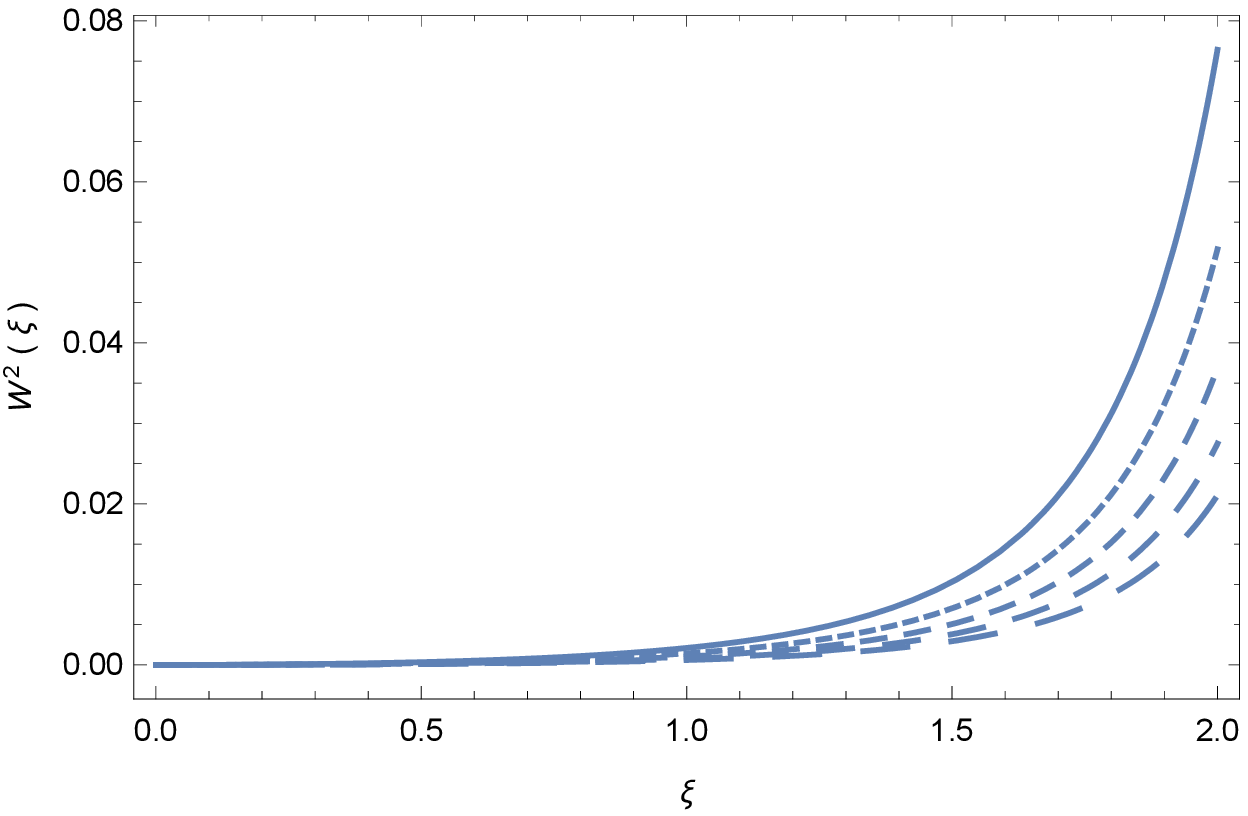}
 \caption{Variation of $W^2(\xi)$ for the cosmic string configuration in the presence of an exponential potential $V(\Phi)=e^{-\Phi}$  for $\Phi \left(0\right)=1$, $W\left(0\right)=10^{-3}$, and for different values of $\Phi_0'$: $\Phi_0'=0.010$ (solid curve), $\Phi_0'=0.012$ (dotted curve), $\Phi_0'=0.014$ (short dashed curve), $\Phi_0'=0.016$ (dashed curve), and $\Phi_0'=0.018$ (long dashed curve), respectively. }
 \label{fig5}
\end{figure}

The variation of  $W^2(\xi)$ is represented in Fig.~\ref{fig5}.
For the adopted set of initial values the behavior of the scalar field and of the metric is very similar to the constant potential case. The scalar field is a monotonically decreasing function of $\xi$, and it reaches the value zero at a greater distance from the string $r=0$ circular line than in the case of the constant potential. The behavior of the field is strongly dependent on the initial conditions. The circular radius $W^2(\xi)$ is a monotonically increasing function of $\xi$, and it is defined properly for $r=0$. However, it becomes singular at a finite distance from the circular line of the string type object, tending to infinity at a finite $\xi$. For distances in the range $\xi \in (0,1)$, or $r\in (0,\sqrt{2/\lambda V_0})$, $W^2(\xi)$ is practically a constant, and its behavior is basically independent on the initial conditions of the scalar field. For the exponential potential the energy density of the string can be obtained generally as a function of the scalar field in the form
\bea
\beta ^2\sigma (\phi)=- \frac{3 }{4 \sqrt{\phi}}\left[C+2 \sqrt{\pi } \beta ^2 \sqrt{\lambda } V_0
   \text{erf}\left(\sqrt{\phi\lambda }\right)\right]
		\nn \\
  - \frac{2 \beta ^2 V_0
   e^{-\lambda  \phi }}{4 \phi}
 \left[\phi  (\lambda  \phi +\lambda +2)+6\right].
\eea

%\begin{widetext}
%\be
%\beta ^2\sigma (\phi)=-\frac{3 \sqrt{\phi } \left[C+2 \sqrt{\pi } \beta ^2 \sqrt{\lambda } V_0
%   \text{erf}\left(\sqrt{\phi\lambda }\right)\right]+2 \beta ^2 V_0
%   e^{-\lambda  \phi } \left[\phi  (\lambda  \phi +\lambda +2)+6\right]}{4 \phi }.
%\ee
%\end{widetext}

However, since the numerical solutions for $\phi(\xi)$ and $W(\xi)$ are known, it is more convenient to obtain $\sigma (\xi)$ from the equation
\be
\beta ^2\kappa ^2 \sigma (x)=-\frac{1+\phi}{W}\left(\frac{d^2W}{d\xi ^2}+\frac{1}{1+\phi}\frac{dW}{d\xi}\frac{d\phi}{d\xi}\right).
\ee

\begin{figure}[tbp]
 \centering
 \includegraphics[scale=0.65]{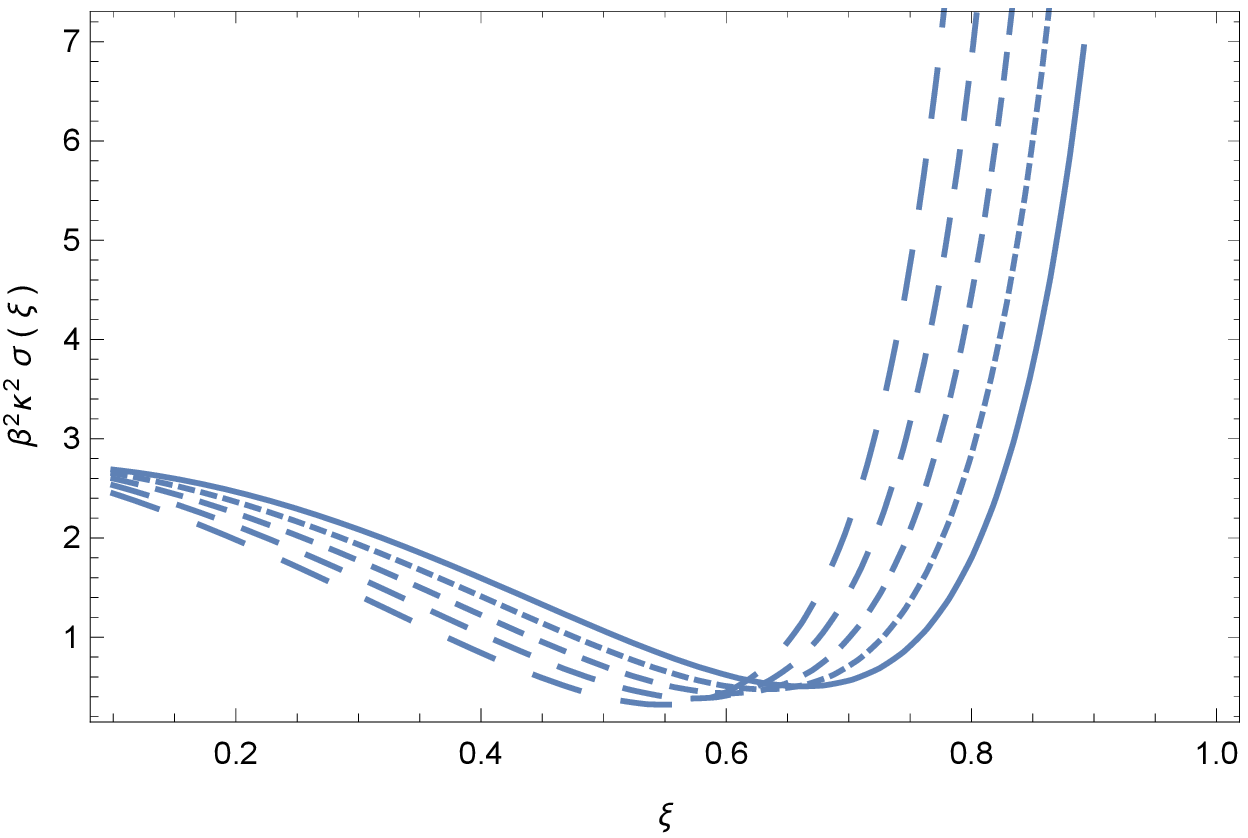}
 \caption{Variation of the energy density $\kappa ^2\sigma (\xi)$ of the cosmic string configuration in the presence of an exponential potential $V(\Phi)=e^{-\Phi}$ for $\Phi \left(0\right)=1$, and for different values of $\Phi_0'$: $\Phi_0'=-0.20$ (solid curve), $\Phi_0'=-0.30$ (dotted curve), $\phi_0'=-0.40$ (short dashed curve), $\Phi_0'=-0.50$ (dashed curve), and $\Phi_0'=-0.60$ (long dashed curve), respectively. }
 \label{fig6}
\end{figure}

The variation of $\sigma$ as a function of $\xi$  is represented in Fig.~\ref{fig6}.
In order to obtain positive energy densities the initial values of $\Phi_0'$ must be negative. There is a significant difference between the behavior of the energy density $\sigma $ as compared to the constant potential case. The energy density initially decreases for increasing values of the radial coordinate, but for $\xi >\xi _{cr}$, the energy density begins to increase, and tends to infinity. Thus the string-like object experiences a singularity at large distances from its circular line $r=0$.

In the first order of approximation, and after rescaling the variable $\phi$,  the integrand in Eq.~(\ref{xiexp}) can be approximated as
\bea
&&\frac{\Phi ^{-3/4}d\Phi }{\sqrt{C+2\sqrt{\pi }%
\text{erf}\left( \sqrt{ \Phi }\right) +2e^{- \Phi }/\sqrt{\Phi }}}\approx     \sqrt{2\Phi }-\frac{C \Phi }{4 \sqrt{2}}
	\nn \\
&&+\frac{\left(3 C^2-16\right)
   \Phi ^{3/2}}{48 \sqrt{2}}+\frac{C \left(48-5 C^2\right) \Phi ^2}{256
   \sqrt{2}}+O\left(\Phi ^{9/4}\right),\nonumber\\
\eea
giving
\bea
\hspace{-0.4cm}\xi +C_0\approx \lambda^{-1/4}\Bigg[ \sqrt{2\Phi }-\frac{C \Phi }{4 \sqrt{2}}+\frac{\left(3 C^2-16\right)
   \Phi ^{3/2}}{48 \sqrt{2}}
	\nn \\
 \hspace{-0.4cm}  +\frac{C \left(48-5 C^2\right) \Phi ^2}{256
   \sqrt{2}}+O\left(\Phi ^{9/4}\right)\Bigg].
\eea
However, this equation is not particularly useful in the study of the behavior of the string models with exponential potential. On the other hand at infinity, we obtain
\be
\frac{\Phi ^{-3/4}d\Phi }{\sqrt{C+2\sqrt{\pi }%
\text{erf}\left( \sqrt{ \Phi }\right) +2e^{- \Phi }/\sqrt{\Phi }}}\approx \frac{1}{\sqrt{C+2 \sqrt{\pi }} \Phi^{3/4}},
\ee
which provides
\be
\xi +C_0\approx \lambda ^{-1/4}\frac{4 \sqrt[4]{\Phi}}{\sqrt{C+2 \sqrt{\pi }}},
\ee
and
\be
\Phi (\xi)\approx  \frac{\lambda}{256} \left(C+2 \sqrt{\pi }\right)^2
   (C_0+\xi )^4,
\ee
respectively.

Within the framework of this approximation the circular radius $W$ is given by the differential equation
\be
\frac{1}{W}\frac{dW}{d\xi}=-\frac{64 e^{-\frac{1}{256} \left(C+2 \sqrt{\pi }\right)^2 \lambda  \left(C_0+\xi
   \right)^4}}{\left(C+2 \sqrt{\pi }\right)^2 \lambda  \left(C_0+\xi
   \right)^3}-\frac{3}{C_0+\xi },
\ee
with the general solution given by
\bea
\hspace{-0.8cm}&W(\xi )=\frac{1}{\left( C_{0}+\xi \right) ^{3}} \exp \Bigg\{2\Bigg[
\frac{16e^{-\frac{1}{256}\left( C+2\sqrt{\pi }\right)
^{2}\lambda \left( C_{0}+\xi \right) ^{4}}}{\left( C_{0}+\xi \right) ^{2}\left( C+2\sqrt{\pi }\right) ^{2}\lambda}
	\nn \\
\hspace{-0.8cm}&+\frac{\sqrt{\pi }}{{\left( C+2\sqrt{\pi }\right) \sqrt{\lambda }}}
	\text{erf}\left(
\frac{1}{16}\left( C+2\sqrt{\pi }\right) \sqrt{\lambda }\left( C_{0}+\xi
\right) ^{2}\right)
\Bigg] \Bigg\}.
\eea

Even in this approximation the full analysis of the behavior of the cosmic string configuration in hybrid metric-Palatini gravity in the presence of an exponential type potential can only be done using numerical methods.
%\begin{widetext}
%\begin{equation}
%W(\xi )=\frac{1}{\left( C_{0}+\xi \right) ^{3}}\exp \left\{ \frac{2\left[
%\sqrt{\pi }\left( C+2\sqrt{\pi }\right) \sqrt{\lambda }
%\text{erf}\left(
%\frac{1}{16}\left( C+2\sqrt{\pi }\right) \sqrt{\lambda }\left( C_{0}+\xi
%\right) ^{2}\right)
%+\frac{16e^{-\frac{1}{256}\left( C+2\sqrt{\pi }\right)
%^{2}\lambda \left( C_{0}+\xi \right) ^{4}}}{\left( C_{0}+\xi \right) ^{2}}%
%\right] }{\left( C+2\sqrt{\pi }\right) ^{2}\lambda }\right\} .
%\end{equation}
%\end{widetext}

%%%%%%%%%%%%%%%%%%%%%%%%%%%%%%%%%%%%%%%%%%%%%%%%%%%%%%%%%%%%%
\subsection{Higgs-type potential}
%%%%%%%%%%%%%%%%%%%%%%%%%%%%%%%%%%%%%%%%%%%%%%%%%%%%%%%%%%%%%

Next we consider the case when the scalar field potential is of the Higgs-type, given by
\be
V(\phi) =\pm \frac{\bar{\mu}^2}{2}\phi ^2+\frac{\nu }{4}\phi ^4,
\ee
where $\bar{\mu}^2$  and $\nu$ are constants. In the following we will investigate only the case with $\bar{\mu}^2<0$, that is, we will adopt the minus sign in the definition of the potential. By following the standard approach in elementary particle physics, we assume that the constant $\bar{\mu}^2$ is related to the mass of the scalar field particle as $m_{\phi}^2= 2\xi v^2 = 2\bar{\mu}^2$, where $v^2 = \bar{\mu}^2/\xi $ gives the
minimum value of the potential. The Higgs self-coupling constant $\nu $ can be obtained, in the case of strong interactions,  from the determination of the mass of the Higgs boson in laboratory experiments, and its numerical value is of the order of $\nu  \approx 1/8$ \cite{Higgs}. By rescaling the radial coordinate and the scalar field according to
\be
r=\sqrt{2}\bar{\mu} \xi, \qquad  \phi=\frac{\Phi}{\left(\nu\bar{\mu}\right)^{1/3}},
\ee
then Eq.~(\ref{xi}) provides the profile of the scalar field in the following form
\be
\frac{d^2\Phi}{d\xi^2}-\frac{3}{4\phi}\left(\frac{d\Phi}{d\xi}\right)^2-\Phi ^2+\Phi^4=0.
\ee
The general solution of this equation is given in a closed form  by
\be
\xi +C_0=\int \frac{1}{\Phi ^{3/4} \sqrt{C+\frac{2}{21} \left(7-3 \Phi ^2\right) \Phi
   ^{3/2}}} \, d\Phi.
\ee
However, this solution cannot be expressed in an analytical form in terms of known functions. In the first order approximation, we obtain
\be
\xi +C_0\approx \frac{4 \sqrt[4]{\Phi }}{\sqrt{C}}-\frac{4 \Phi ^{7/4}}{21
   C^{3/2}}+O\left(\Phi ^{9/4}\right),
\ee
but this representation is not particularly useful from the point of view of concrete calculations.

\begin{figure}[tbp]
 \centering
 \includegraphics[scale=0.65]{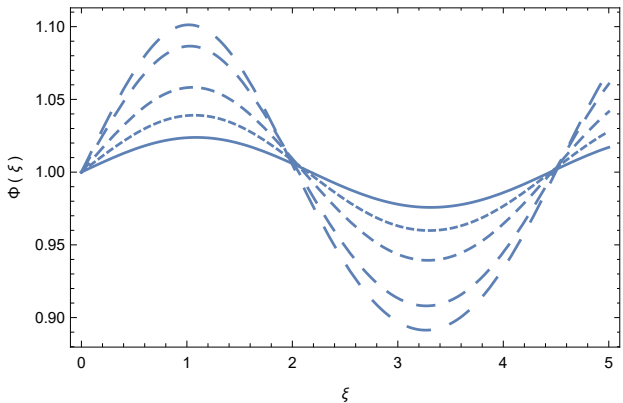}
 \caption{Variation of the scalar field for the cosmic string configuration in the presence of a Higgs-type potential $V(\Phi)=-\Phi^2+\Phi^4$ for $\Phi (0)=\Phi_0=1$, and for different values of $\Phi_0'$: $\Phi_0'=0.034$ (solid curve), $\Phi_0'=0.056$ (dotted curve), $\Phi_0'=0.084$ (short dashed curve), $\Phi_0'=0.126$ (dashed curve), and $\Phi_0'=0.148$ (long dashed curve), respectively.}
 \label{fig7}
\end{figure}

The variation of the scalar field with Higgs potential supporting a string configuration in hybrid metric-Palatini gravity is represented in Fig.~\ref{fig7}.
There is a significant qualitative difference between this string model and the constant, simple power law or exponential potentials. Note that the scalar field for the Higgs-type potential shows a basically periodic structure, changing between successive maxima and minima. There are singularities in the field. Its behavior is strongly affected by the initial conditions on the circular line, and the field extends to infinity.

\begin{figure}[tbp]
 \centering
 \includegraphics[scale=0.65]{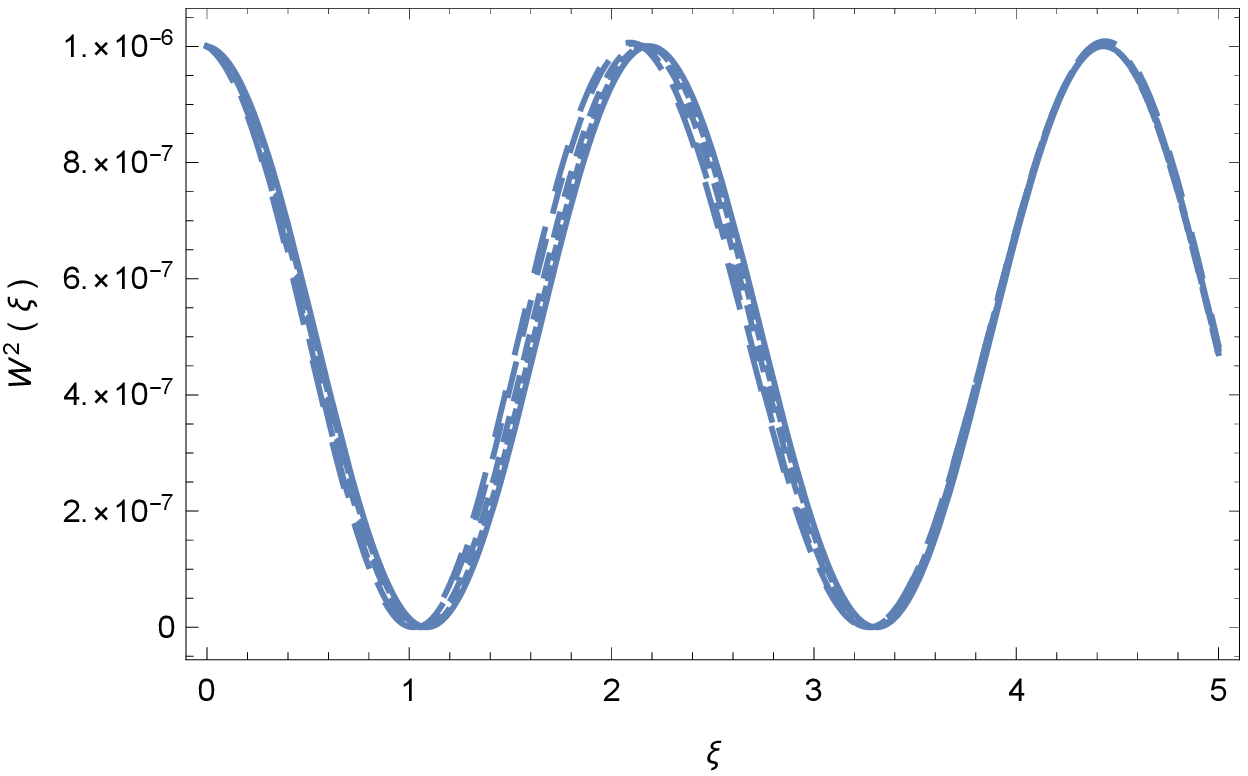}
 \caption{Variation of $W^2(\xi)$ for the cosmic string configuration in the presence of a Higgs type potential $V(\Phi)=-\Phi^2+\Phi^4$  for $\Phi \left(0\right)=1$, $W\left(0\right)=10^{-3}$, and for different values of $\Phi_0'$: $\Phi_0'=0.034$ (solid curve), $\Phi_0'=0.056$ (dotted curve), $\Phi_0'=0.084$ (short dashed curve), $\Phi_0'=0.126$ (dashed curve), and $\Phi_0'=0.148$ (long dashed curve), respectively.}
 \label{fig8}
\end{figure}

The variation of  $W^2(\xi)$ in the presence of a Higgs potential is represented in Fig.~\ref{fig8}.
The same oscillatory pattern can also be observed in the case of the circular rdaius $W^2$. However, there is a difference in the phase of these to quantities. When the field reaches its maximum at $\xi \approx 1$, the metric tends to zero, $W^2(1)\approx 0$. Then, while the scalar field decreases,  $W^2(\xi)$ increases, reaching its maximum at the minimum of the field, corresponding to $\xi \approx 2$. This pattern is repeated up to infinity. On the other hand, as one see Fig.~\ref{fig8}, the actual physical solution may be considered as being defined between to adjacent zeros of $W(\xi)$, which
are centers on the $\left(\xi, \theta\right)$ surface. Hence the obtained string configuration may represent a closed world in these two directions.

\begin{figure}[tbp]
 \centering
 \includegraphics[scale=0.65]{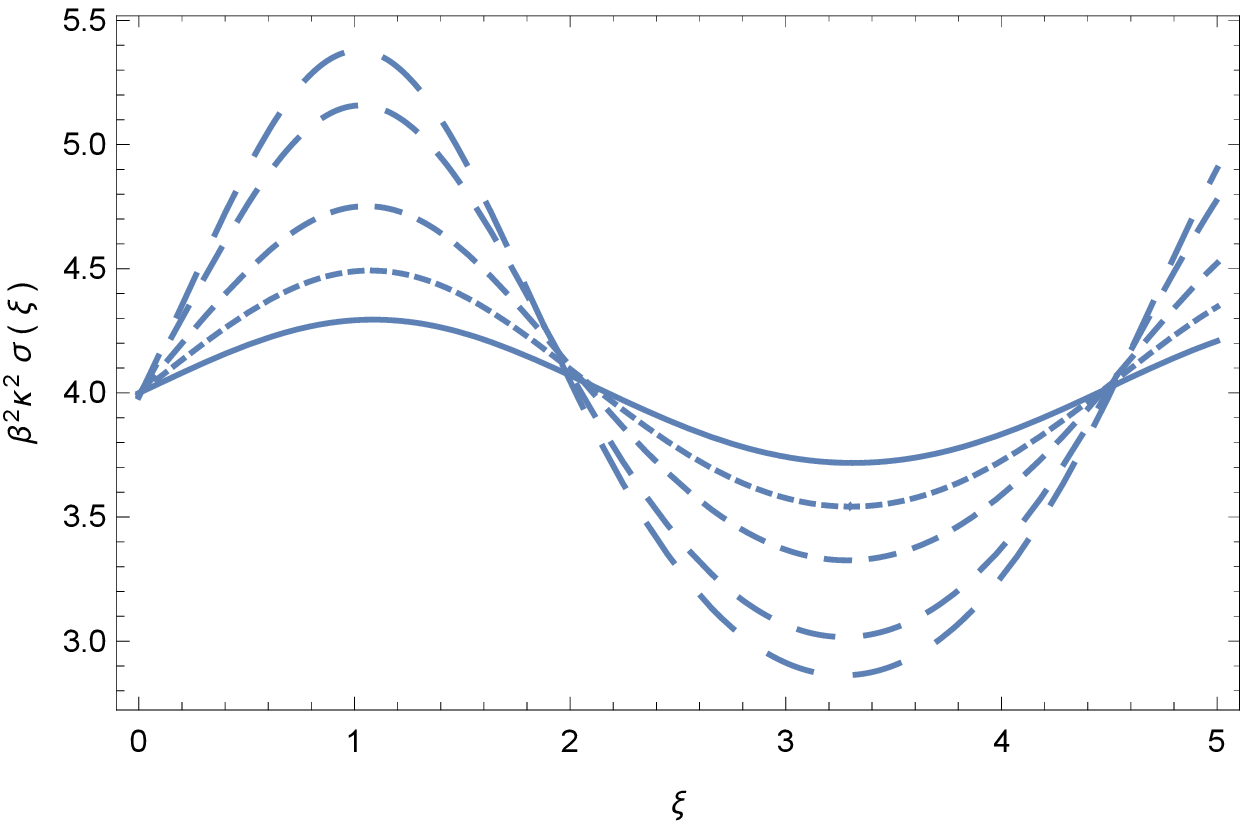}
 \caption{Variation of the string tension $\kappa ^2\sigma (\xi)$ of the cosmic string configuration in the presence of a Higgs type potential $V(\Phi)=-\Phi^2+\Phi^4$  for $\Phi \left(0\right)=1$, $W\left(0\right)=10^{-3}$, and for different values of $\Phi_0'$: $\Phi_0'=0.034$ (solid curve), $\Phi_0'=0.056$ (dotted curve), $\Phi_0'=0.084$ (short dashed curve), $\Phi_0'=0.126$ (dashed curve), and $\Phi_0'=0.148$ (long dashed curve), respectively.}
 \label{fig9}
\end{figure}

The variation of the string tension with respect to the radial coordinate in the presence of the Higgs potential is depicted in Fig.~\ref{fig9}.
The variation of  $\sigma$ is in phase with that of the scalar field, and both quantities reach their maxima and minima at the same position. The string tension also has an oscillatory behavior, which is a general property of all physical and geometrical parameters of the string configurations supported by scalar fields with a Higgs-type potential.

%%%%%%%%%%%%%%%%%%%%%%%%%%%%%%%%%%%%%%%%%%%%%%%%%%%%%%%%%%%%%
\section{Discussions and final remarks}\label{sec:V}
%%%%%%%%%%%%%%%%%%%%%%%%%%%%%%%%%%%%%%%%%%%%%%%%%%%%%%%%%%%%%

In the present paper we have investigated string-type solutions in hybrid metric-Palatini gravity, which is an extension of general relativity that combines the metric and Palatini formalisms. From a
theoretical point of view the main advantage of the hybrid metric-Palatini theory is that it is a viable theory of gravity that includes elements of both formalisms. A main success of the theory is the possibility of generating long-range forces that pass the classical local tests at the Solar System level of
gravity. Another important advantage of the theory is that it admits an equivalent scalar-tensor representation, which greatly simplifies the analysis of the field equations, and the construction of their solutions. In this work, we have explored local gauge string solutions with a phenomenological energy momentum tensor, as prescribed in \cite{Vilenkin}.

In our present investigation of the cosmic string-type solutions in the context of the hybrid metric-Palatini gravitational theory we have adopted, after several simplifications, the metric (\ref{metrn}), of the form similar to the metric (\ref{stringmet}), with $W^2(r)=\left[1-\mu (r)/2\pi\right]^2r^2$. By adopting the scalar-tensor representation, the gravitational field equations can be formulated in terms of the circular radius $W$, the scalar field and its derivatives, and the string tension, respectively. The model also contains as an essential ingredient the scalar field potential $V$. Generally, for the case of an arbitrary potential, the general solution of the field equations can be obtained in a closed form. For at least three particular choices of the scalar field potential the solution of the field equations can be expressed in an exact (and simple) analytic form. It is important to point out that for all considered solutions $W^2(0)\neq 0$, and therefore one cannot define an axis for these solutions \cite{Steph}. That's why they do not represent proper string solutions of the gravitational field equations, and their properties show significant differences with respect to the standard string solutions of the other geometric gravitational field theories.

For some types of potentials the string solutions in the hybrid theory have some important distinctive features as compared to the other string models. First of all, the behavior of the solutions is strongly dependent on the initial conditions of the scalar field and of its derivative for $r=0$. These initial conditions are rather arbitrary, since a large number of such field configurations can be constructed. Depending on the initial conditions for the field we obtain two distinct classes of solutions. The first class consists of the solutions that become singular for a finite value of the rescaled radial coordinate $\xi$. For example, in the string solution with $V=0$, if $\phi _0>0$ and $\phi _0'<0$, then $C_0=\phi _0/\phi'_0>0$, and the metric and the string tension are becoming singular (infinite) at $\xi _{\infty}=4\phi _0\phi'_0$. In this case the scalar field vanishes for $\xi _{\infty}$.
A completely different situation arises if both $\phi _0$ and $\phi _0'$ are positive. In this case both the string tension and the metric tend to zero at infinity, with the scalar field becoming singular for $\xi \rightarrow \infty$.

The exact solution of the field equations corresponding to the $V=V_0\phi ^{3/4}$ corresponds to the second class. The requirement of the positivity of the metric tensor components for $r=0$ imposes the condition that the metric tensor components and the string tension are decreasing functions of the radial coordinate $r$, since both $\phi _0$ and $\phi _0'$ must be positive. As a consequence, it is the scalar field that diverges at infinity.    However, in these cases one could also define a string radius by introducing an effective cut-off length $\xi _{\rm co}$ for the metric and scalar field, which would allow to construct finite string configurations, with finite values of the scalar field, string tension and metric tensor components. But in this case the definition of $\xi _{\rm co}$ is either arbitrary, or based on some empirical considerations, such as the consistency with observational data.

In the case of the exponential and Higgs type scalar field potentials, one can use the condition of the vanishing of the string tension $\sigma (r)$ to define a radius of the string. This can be done only numerically, and the numerical value of the string radius is strongly dependent on the initial conditions of the scalar field and its derivative at $r=0$. For example, in the case of the exponential potential, the string tension vanishes at $\xi _{\rm cr}\approx 0.6$, which would allow to define a string radius $R_s$ of the order of magnitude $R_s\approx 0.6 \sqrt{2/V_0\lambda}\,\xi _{\rm cr}$. Generally, the exponential potential $\sigma $ reaches its zero/minimum value for finite values of the scalar field potential, and of the metric tensor components, and hence the condition $\sigma \approx 0$ may define the string radius for this type of scalar field potential.

The case of the Higgs potential is quite interesting. The string tension does not vanish for any value of the radial coordinate, and it reaches its minimum value for $\xi \approx 3.5$. For this value of $\xi$ the scalar field is at its minimum, while the circular radius $W^2(\xi)$ is singular, and tends to zero. Alternatively, one could define the string radius as corresponding to the first zero of the metric tensor component $W^2(\xi)$ ($\xi \approx 1$), with the scalar field and the string tension reaching their first maxima. One can also introduce a cut-off radius for these potentials, with its value being determined from the confrontation of the theoretical predictions with observations. For example, one can construct a string configuration by introducing a cutoff radius of $W^@(\xi)$ before reaching the singularity. In the case of the Higgs potential we may choose $\xi_s\approx 0.90$, and then we match the solution either with the cosmological background, or with another (standard general relativistic) string solution. In this way we obtain a well-behaved string like structure in the interval $\xi \in [0, \xi _s]$. Another possibility is to consider as the physical string solution the interval $\xi _0\leq \xi \leq \xi _1$, where $\xi _0$ and $\xi _1$ are the two zeros of $W^2(\xi)$. In this configuration the string has two singular points at its extremities.

In Section~\ref{reg} we have presented, following \cite{Bronstring}, the standard regularity and asymptotic conditions required for string-like solutions of the gravitational field equations. They are regular on the string axis, regular at infinity, and possess a finite mass per unit length. Generally, all our solutions satisfy the regularity condition on the circular line, with the circular radius $W^2$ taking finite values for $\xi =0$. $W^2(0)$ is determined by the initial conditions satisfied by the scalar field. On the other hand, very few of our solutions satisfy the asymptotic conditions at infinity, with either $W^2(\xi)$, the scalar field $\phi$, or total mass being singular (zero or infinite) at infinity. From a physical point of view we explain this result as due to two factors: a) in our investigations we have used the scalar-tensor representation of the hybrid metric-Palatini gravity theory. As pointed out in  \cite{Bronstring}, string solutions in scalar-tensor  field models do not satisfy the asymptotic conditions, and this may also be the case for the scalar-tensor type theory we have used in our investigations,  and b) the hybrid metric-Palatini gravity theory naturally explains the recent acceleration of the Universe.  Therefore it represents a viable alternative to the dark energy paradigm, and to the standard $\Lambda$CDM cosmological model. Thus, in hybrid metric-Palatini gravity an effective cosmological constant  naturally appears. Far from the gravitating sources, in spherical symmetry the scalar field behaves as
\be
\phi (r)\approx \phi _0^{(\rm vac)}+\frac{2G\phi _0^{(b)}M}{3r}e^{-m_{\phi}r},
\ee
with the effective mass $m_{\phi}$ given by
\begin{equation*}
m_{\phi}=\frac{1}{3}\left[2V-V_{,\phi}-\phi(1+\phi)V_{,\phi \phi}\right]\big|_{\phi =\phi _0^{(b)}},
\end{equation*}
where $\phi _0^{(b)}$ is the background value of the scalar field. For the metric perturbations we obtain  \cite{Harko:2011nh, Capozziello:2015lza},
\be\label{minka}
h_{00}^{(2)}=\frac{2G_{\rm eff}M}{r}+\frac{V_0^{(\rm vac)}r^2}{6\left(1+\phi _0^{(b)}\right)},
 \ee
where $V_0^{(b)}=V\left(\phi _0^{(b)}\right)$ is the background value of the scalar field potential, and $G_{\rm eff}$ is the effective gravitational constant. It is obvious that $h_{00}^{(2)}$ diverges as $r\rightarrow \infty$. This implies that the vacuum at infinity is of the de Sitter type, and that generally the solutions of the field equations in hybrid metric-Palatini gravity are not asymptotically flat. Hence the divergence at infinity of our solutions is an indicator of the presence of a vacuum energy (an effective cosmological constant) at infinity. This also indicates that our string solutions are of cosmological type.

We would also like to point out that the singular behavior of the solutions, being of cosmological nature, happens at the ``cosmological infinity'', at distances from the circular line having the order of magnitude of the Hubble radius. These distances are much larger than those we may associate to the ``astrophysical infinity''. Hence we may assume that in the range of the distances relevant astrophysically (up to the galaxy cluster level, for example), one can construct regular string solutions, by introducing a cut-off radius for the string.

Since the hybrid metric-Palatini gravity theory is intrinsically ``cosmological'', with the vacuum at infinity being of de Sitter type, in order to construct physically realistic solutions a cutoff procedure must be introduced, in order to avoid any pathological behavior of the geometrical and physical quantities.
If we assume that the cosmic strings extend indefinitely with the increasing distance from the central circular line, the resulting spacetime is
not asymptotically flat, but of de Sitter type. This is due to the presence of the (effective) cosmological constant in the Universe.  There are two possibilities to estimate the upper bound for the cutoff of the strings. The first one, we call cosmological cutoff, is based on the idea  to define the string radius as the distance from the circular line at which the decreasing density profile of the string becomes smaller than the average energy density of the Universe. More specifically, we define the cosmological cutoff radius $\xi _s$ of the string as the finite distance from the $r=0$ circular line that satisfies the following two conditions: a) the scalar field takes its background value $\phi \left(\xi _s\right)=\phi _0$, and b) the string tension satisfies the condition $\sigma \left(\xi _s\right)\leq \rho_{\rm univ}$, where $\rho_{\rm univ}$ is the mean energy density of the Universe.
The mean density of the Universe as well as  the numerical value of the effective cosmological constant can be also expressed by using  the density parameters $\Omega =\rho_{\rm univ}/\rho_{\rm crit}$ and $\Omega _{\Lambda}=\Lambda c^2/3H_0^2$, where $H_0$ is the Hubble constant, and $\rho_{\rm crit}=3H_0^2/8\pi G$ is the critical density of the Universe.

We can obtain a simple qualitative estimate of the extension of a cosmic string from the following simple computation. By assuming that the (effective) cosmological constant has a numerical value of the order of $\Lambda =3 \times  10^{-56}\; {\rm cm}^{-2}$ \cite{Planck}, at a distance from the circular line of the order of $r = 100$ kpc, for a supermassive cosmic string with mass $M = 10^{10}M_{\odot}$, the quantities $GM/r=4.94\times 10^{-9}$ and $\Lambda r^2=2.7\times 10^{-9}$, appearing in Eq.~(\ref{minka}),  are roughly of the same order of magnitude. For radial distances greater than 100 kpc the effects of the cosmological constant become dominant. This example shows that even in the case of an extremely heavy cosmic string the infinities of the theory are of astrophysical type, of the order of hundreds of kiloparsecs, much smaller than the ``cosmological infinities'' of the order of the Hubble radius of the Universe, which is around 3000 megaparsecs. On the other hand, for the case of a small mass cosmic string with $M=10^{-20}M_{\odot}$, the gravitational potential energy and cosmological expansion energy are of the same order of magnitude for $r=0.11$ kpc only.

As an example of the application of the cosmological cutoff procedure we will consider the case of the simplest string solution, corresponding to $V=0$,  and given by Eqs.~(\ref{ss1}), (\ref{ss2}), and (\ref{ss3}), respectively. Hence, in order to define a proper radius of the string, we require that at some finite radius $\xi _s$ the scalar field reaches its background value, $\phi \left(\xi _s\right)=\phi _0^{(b)}$, which gives for $\xi _s$ the expression
\be
\xi _s=4C_0+\left(\frac{\phi _0^{(b)}}{C}\right)^{1/4}.
\ee
The condition of the equality of the string tension with the critical density of the Universe, $\left|\sigma \left(\xi_s\right)\right|=\rho _{\rm crit}=3H_0^2/8\pi G$,  gives for the integration constant $C$ the expression $C=\left(H_0^2/32\pi G\right)^2\phi _0^{(b)}$, while for $W^2\left(\xi _s\right)$ we obtain $W^2\left(\xi _s\right)=w_0^2C^{3/2}\left(\phi _0^{(b)}\right)^{-3/2}$. The metric is not asymptotically flat, and $\xi >\xi _s$ the cosmological background dominates the string dynamics.

A second possibility for defining a cutoff radius for the string is to match the hybrid metric-Palatini gravity string-like solutions with another, exterior string solution. If  at radii exceeding some radius specific $r_s$ the effects of the scalar field may become negligibly small, and hybrid metric-Palatini gravity reduces to standard general relativity,  then the present string solutions may go smoothly into a general relativistic cylindrically symmetric solution of the Einstein field equations. For example, by assuming that in a certain limit the string-like metrics obtained in the present study reduce to the metric (\ref{Vil1}), we can again define a finite string radius by imposing the conditions $W^2\left(\xi _s\right)=\left(1-8\pi G\mu\right)\beta ^2 \xi _s^2$, and $\sigma \left(\xi _s\right)=\mu$, where the mass per unit length of the string must satisfy the condition $\mu ={\rm constant}$ for all $\xi \geq \xi _s$.

In the case of the zero potential the matching of the two metrics gives the string radius as
\be
\xi _s=4C_0+\sqrt{\frac{12}{\mu}},
\ee
with $\mu $ obtained as a solution of the algebraic equation
\be
w_0^2\left(\frac{\mu }{12}\right)=\left(1-8\pi G\mu\right)\beta ^2\left(4C_0+\sqrt{\frac{12}{\mu}}\right)^2.
\ee
For the scalar field at the string boundary we obtain the expression
\be
\phi \left(\xi _s\right)=\frac{144C}{\mu ^2}.
\ee

On the other hand, for some scalar field potential types one could define the boundary of the string as the point where the energy density of the string vanishes. But even in this case one should match the string solution either with the cosmological background, or with an exterior string solution.

An important geometrical quantity, the angular deficit $\Delta \theta$ in the cylindrical symmetry, due to the presence of the string, is given by \cite{Verbin}
\be\label{def}
\Delta \theta =2\pi \left[1-\lim_{r\rightarrow \infty}W'\left(r\right)\right],
\ee
In the first order of approximation, and for strings with finite extension,  we may replace $W'\left(\infty\right)$ with $W'\left(R_s\right)$
in Eq. (\ref{def}), where $R_s$ is the string radius, thus obtaining
\be\label{def1}
\Delta \theta \approx 2\pi \left[1-W'\left(R_s\right)\right].
\ee
In the variable $\xi$ the angular deficit can be represented as
\be
\Delta \theta \approx 2\pi \left[1-\frac{1}{\beta}W'\left(\xi_s\right)\right].
\ee
Since the variation of $W$ depends on the initial conditions of the scalar field on the $r=0$ circular line, the string-like geometries obtained in the present study allow a very large range of deficit angles, which significantly impact the geometry of the spacetime near the string. For the solutions with $\lim_{\xi \rightarrow \infty} W(\xi)\rightarrow 0$, generally also $\lim_{\xi \rightarrow \infty} W'(\xi)\rightarrow 0$, like, for example, in the zero potential case with $\phi _0$ and $\phi _0'$ positive. In this case we obtain $\Delta \theta \approx 2\pi$. In the opposite limit of $\lim_{\xi \rightarrow \infty} W'(\xi)\rightarrow \infty$, the deficit angle is formally infinite. We can still define a finite deficit angle by introducing a cut-off radius $R_s$ that formally defines the radius of the string.  For the exponential and Higgs potentials one can define explicitly the radius of the string-like objects, which also allows the explicit estimation of the deficit angle.

Cosmic strings have a number of very intriguing properties. For example, as suggested by Witten \cite{Witten}, strings behave like superconducting wires. Hence they can interact with external cosmic electromagnetic fields,  and  as they move through cosmic magnetic fields they can develop electric currents. Therefore, short electromagnetic and highly beamed
bursts can be emitted from  some peculiar points (cusps), located on small string segments, where the velocity approaches the speed of
light \cite{C1, C2, C3, C4,C5}. Hence the cusp is a powerful source of electromagnetic radiation that may produce a jet of accelerated particles that may play an important role  in many astrophysical phenomena, such as Gamma Ray Bursts and afterglow emissions, respectively. It would be interesting to consider superconducting strings in the framework of modified theories of gravity, and in particular in hybrid metric-Palatini gravity. Such a study would offer some possibilities between discriminating standard cosmic strings from string-like structures that appear in modified theories of gravity.

Another important physical effect that could, at least in principle, discriminate between standard general relativistic cosmic strings, cosmic strings in modified gravity,  and other filamentary matter distributions is gravitational lensing.  According to the standard general relativistic string scenario, the curvature of the spacetime is not changed by a vacuum string. However, the topology of the spacetime is modified \cite{Gott}. Hence photon beams are not bent by a cosmic string. But if two light rays travel on the different sides of the string, the presence of the specific conical structure of the spacetime geometry determines their later convergence at the same point of observation \cite{Bozza}. Hence, for a cosmic string located between a terrestrial observer and a distant cosmological source, the observer will detect two images of the light emitting source, separated by an angle $\delta \theta =8 \mu _s \sin \alpha D_{\rm LS}/D_{\rm OS}$, where by $\mu _s$ we have denoted the linear mass density of the string, $\alpha$ represents the angle between the observer-source direction and the string, while $D_{\rm OS}$ and  $D_{\rm LS}$  represent the distance between the observer and the source, and the lens and the source, respectively. Hence it follows that for the case of the standard general relativistic conical string, due to the string presence, the two images formed are identical to the original source, without any distortion or amplification \cite{Bozza}. This effect is very different from the gravitational lensing by gas filaments, which show a very different image structure, formed from one or three elongated images \cite{Bozza}. Hence the lensing properties of the string solutions in hybrid metric-Palatini gravity obtained in the present study could help in discriminating between these string solutions and the corresponding solutions obtained in standard string theory, or other modified gravity models. The systematic study of the lensing properties of hybrid metric-Palatini gravity strings, and their observational implications, will be considered in a future work.

The in-depth investigation of modified theories of gravity and of their astrophysical and cosmological implications is a major field of study in present day theoretical physics. Despite the fact that from the observational point of view cosmic strings are still elusive astrophysical and cosmological objects, the investigation of their theoretical properties may lead to a better understanding of the theoretical structure of modified gravity. In the present paper
we have provided some basic theoretical tools that would enable the in-depth investigation of the properties of the cosmic strings in the hybrid metric-Palatini theory of gravity, and of their astrophysical and cosmological implications.

%%%%%%%%%%%%%%%%%%%%%%%%%%%%%%%%%%%%%%%%%%%%%%%%%%%%%%%%%%%%%
\acknowledgments
%%%%%%%%%%%%%%%%%%%%%%%%%%%%%%%%%%%%%%%%%%%%%%%%%%%%%%%%%%%%%

FSNL acknowledges support from the Funda\c{c}\~{a}o para a Ci\^{e}ncia e a Tecnologia (FCT) Scientific Employment Stimulus contract with reference CEECIND/04057/2017, and the research grants No. UID/FIS/04434/2019 and No. PTDC/FIS-OUT/29048/2017.
%%%%%%%%%%%%%%%%%%%%%%%%%%%%%%%%%%%%%%%%%%%%%%%%%%%%%%%%%%%%%

%\enlargethispage{50pt}

%\clearpage

%%%%%%%%%%%%%%%%%%%%%%%%%%%%%%%%%%%%%%%%%%%%%%%%%%%%%%%%%%%%%

%%%%%%%%%%%%%%%%%%%%%%%%%%%%%%%%%%%%%%%%%%%%%%%%%%%%%%%%%%%%%
\end{document}